\documentclass[]{article}
\usepackage[]{amsmath}
\usepackage[]{amssymb}
\usepackage[]{amsthm}
\usepackage{geometry}
\usepackage[onehalfspacing]{setspace}
\usepackage{graphicx}
\usepackage{rotating}
\usepackage[numbers, square]{natbib}
\usepackage{color}
\usepackage{url}
\usepackage{authblk}
\usepackage{comment}
\usepackage{tikz}
\usetikzlibrary{positioning}
\DeclareMathOperator*{\argmin}{arg\,min}

\begin{document}
	
\title{Pleiotropy robust methods for multivariable Mendelian randomization}
\author[1]{Andrew J. Grant\thanks{Corresponding author. Email address: andrew.grant@mrc-bsu.cam.ac.uk}}
\author[1,2]{Stephen Burgess}
\affil[1]{\normalsize MRC Biostatistics Unit, University of Cambridge, Cambridge, UK}
\affil[2]{\normalsize Cardiovascular Epidemiology Unit, University of Cambridge, Cambridge, UK}
\date{}
\maketitle

\begin{abstract}
	Mendelian randomization is a powerful tool for inferring the presence, or otherwise, of causal effects from observational data. However, the nature of genetic variants is such that pleiotropy remains a barrier to valid causal effect estimation. There are many options in the literature for pleiotropy robust methods when studying the effects of a single risk factor on an outcome. However, there are few pleiotropy robust methods in the multivariable setting, that is, when there are multiple risk factors of interest. In this paper we introduce three methods which build on common approaches in the univariable setting: MVMR-Robust; MVMR-Median; and MVMR-Lasso. We discuss the properties of each of these methods and examine their performance in comparison to existing approaches in a simulation study. MVMR-Robust is shown to outperform existing outlier robust approaches when there are low levels of pleiotropy. MVMR-Lasso provides the best estimation in terms of mean squared error for moderate to high levels of pleiotropy, and can provide valid inference in a three sample setting. MVMR-Median performs well in terms of estimation across all scenarios considered, and provides valid inference up to a moderate level of pleiotropy. We demonstrate the methods in an applied example looking at the effects of intelligence, education and household income on the risk of Alzheimer's disease.
\end{abstract}

\section{Introduction}
Mendelian randomization is a technique for estimating the causal effect of a risk factor on an outcome using observational data \citep{GDS2003}. It uses genetic variants as instrumental variables and can provide valid causal effect estimation in the presence of unmeasured confounding. Three  assumptions are required in order that a genetic variant is a valid instrument: it must be associated with the risk factor of interest; it must not be associated with any confounder of the risk factor-outcome relationship; and it must be independent of the outcome conditional on the risk factor and confounders \citep{Greenland2000}.

Genetic variants are good candidates for instrumental variables: they are naturally independent of many environmental factors which are common sources of confounding, and mitigate the potential for reverse causation. Furthermore, methods for Mendelian randomization have been developed which allow for combining many instruments in a single analysis, and which can also be used when only summary statistics of the associations between the genetic variants and traits are available \citep{BButterworthThompson2013}. These features allow practitioners to harness publicly available summary data from genome wide association studies (GWAS). Two-sample approaches, where the genetic variant-risk factor and genetic variant-outcome associations are estimated in different samples, open up vast combinations of risk factor-outcome relationships to be studied \citep{PierceBurgess2013}. The major limitation in Mendelian randomization analyses is therefore the potential presence of pleiotropy, which is when genetic variants associate with traits other than the risk factor of interest. If any such trait provides an alternative causal pathway to the outcome not via the risk factor, then the corresponding genetic variants are invalid instruments and causal effect estimates may be biased.

Multivariable Mendelian randomization fits multiple risk factors in a single model \citep{BThompson2015mv}. One motivation for its use is to account for pleiotropy in a univariable analysis via a set of measured covariates. It can be an important sensitivity analysis if there are known biological pathways linking the genetic variants and the outcome. Another motivation is if there are a number of correlated traits with shared genetic predictors which are all hypothesized to have potential causal effects on the outcome. A multivariable model can distinguish between the direct effects of the risk factors on the outcome and the total effects inclusive of mediators \citep{Sanderson2019}. A genetic variant is a valid instrument for multivariable Mendelian randomization if: it is associated with at least one risk factor; it is independent of any confounder of each risk factor-outcome relationship; and it is independent of the outcome conditional on all risk factors and confounders. Causal pathways from a genetic variant to the outcome that do not pass via one or more of the risk factors are referred to as unmeasured pleiotropy (in contrast to measured pleiotropy, where such pathways are entirely account for via the set of risk factors). For the purposes of this paper, we use the word pleiotropy to mean unmeasured pleiotropy.

There are a number methods in the literature for univariable Mendelian randomization (that is, when there is a single risk factor) which are robust to pleiotropy \citep{Slob2020}. Each method provides valid estimation of the causal effect under different sets of assumptions. Although these assumptions are, generally, untestable, an applied analysis will typically employ a range of methods. Consistency of results across various methods which rely on different assumptions gives strength of evidence to the findings \citep{Lawlor2017}. There are, however, few methods for pleiotropy robust multivariable Mendelian randomization. Valid estimation of causal effects, therefore, typically relies on the assumption that all causal pathways between the genetic variants and the outcome are accounted for via the measured risk factors.

In this paper we propose a number of novel approaches to multivariable Mendelian randomization which provide robustness to different forms of pleiotropy. The methods are developed for use with summary level data, and so access to individual level data is not required. We examine the performance of the methods under various pleiotropic settings in a simulation study. We then demonstrate the methods in an applied analysis looking at the effects of intelligence, years of education and household income on the risk of Alzheimer's disease.

\section{Modelling assumptions}
\subsection{Data generating model}
We assume the following model, which is similar to a multivariable version of the one set out by \citet{Bowden20172sample} in the single risk factor case. For individual $i$, let $Y_{i}$ be the outcome, $X_{i1}, \ldots, X_{iK}$ be $K$ risk factors, $G_{i1}, \ldots, G_{ip}$ be $p$ genetic variants and $U_{i}$ represent confounders of the risk factor-outcome relationships. The data generating model is:
\begin{align}
	X_{ik} &= \beta_{X0k} + \sum_{j=1}^{p} \beta_{Xjk} G_{ij} + \gamma_{Xk} U_{i} + v_{Xik} , \quad k = 1, \ldots, K \label{eq:X} \\
	Y_{i} &= \theta_{0} + \sum_{k=1}^{K} \theta_{k} X_{ik} + \sum_{j=1}^{p} \alpha_{j} G_{ij} + \gamma_{Y} U_{i} + v_{Yi}, \label{eq:Y}
\end{align}
where $v_{Xik}$ and $v_{Yi}$ are independent error terms with mean zero. Note that the $v_{Xik}$ are not necessarily independent of each other, and so the risk factors may be correlated via the correlation between these error terms as well as their common association with $U_{i}$. We assume that the genetic variants are independent of each other and independent of $U_{i}$. We further assume that $p > K$ and the $p \times K$ matrix with ($j$, $k$)\textsuperscript{th} element $\beta_{Xjk}$ is of full column rank.

\subsection{Instrument validity and pleiotropy}
\begin{figure}
	\centering
	\begin{tikzpicture}[]
	\node[] (a) {$\vdots$};
	\node[] (b) [right=2.25cm of a] {};
	\node[] (Gj) [left=3cm of a] {$G_{j}$};
	\node[] (X1) [above=0.5cm of a] {$X_{1}$};
	\node[] (XK) [below=0.5cm of a] {$X_{K}$};
	\node[] (Y) [right=3cm of a] {$Y$};
	\node[] (U) [above=2cm of b] {$U$};
	
	\path[-stealth]
	(U) edge node[above] {\footnotesize $\gamma_{X1}$} (X1)
	(U) edge node[above right = 0.8cm] {\footnotesize $\gamma_{XK}$} (XK)
	(U) edge node[above = 0.5cm] {\footnotesize $\gamma_{Y}$} (Y)
	(Gj) edge node[above] {\footnotesize $\beta_{Xj1}$} (X1)
	(Gj) edge node[above] {\footnotesize $\beta_{XjK}$} (XK)
	(Gj) edge[bend right=75] node[below] {\footnotesize $\alpha_{j}$} (Y)
	(X1) edge node[above] {\footnotesize $\theta_{1}$} (Y)
	(XK) edge node[above] {\footnotesize $\theta_{K}$} (Y);
	\end{tikzpicture}
	\caption{Directed acyclic graph showing the relationship between the $j$\textsuperscript{th} genetic variant ($G_{j}$), the risk factors ($X_{1}, \ldots, X_{K}$), confounders ($U$) and the outcome ($Y$).}
	\label{fg:dag1}
\end{figure}
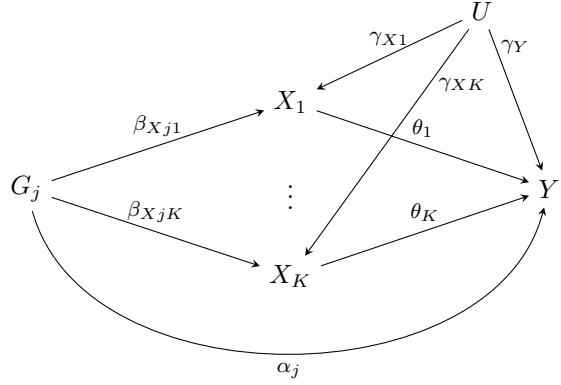

The relationships between a single genetic variant, the risk factors, confounders and outcome in model (\ref{eq:X})--(\ref{eq:Y}) are represented by the directed acyclic graph in Figure \ref{fg:dag1}. For the $j$\textsuperscript{th} genetic variant, pleiotropy is caused by the $\alpha_{j}$ term. Since the model allows for no direct association between the genetic variant and $U_{i}$, $G_{j}$ is a valid instrument if at least one of $\beta_{Xj1}, \ldots, \beta_{XjK}$ are non-zero and $\alpha_{j} = 0$. Note that although the model suggests $\alpha_{j}$ represents direct effects of the genetic variant on the outcome, it may also represent an association via an unmeasured trait.

If $\alpha_{j} = 0$ for all $j$, then there is no pleiotropy and all genetic variants are valid instruments. When not all $\alpha_{j}$'s are zero, we consider two patterns of pleiotropy. The first is referred to as balanced pleiotropy, which is where the $\alpha_{j}$'s are distributed with mean zero. The second is referred to as directional pleiotropy, which is where the $\alpha_{j}$'s are distributed with mean not equal to zero. In each case we assume that the $\alpha_{j}$'s are independent of each other. It may also be that most of the $\alpha_{j}$ are equal to zero but a relatively small number of them are non-zero and possibly large in magnitude. We will refer to these non-zero $\alpha_{j}$'s as outliers.

\subsection{Summary level data}
We denote by $\hat{\beta}_{Xjk}$ and $\hat{\beta}_{Yj}$ the estimates obtained by regressing the $k$\textsuperscript{th} risk factor and outcome, respectively, on the $j$\textsuperscript{th} genetic variant. We have that
\begin{align*}
	\hat{\beta}_{Xjk} &= \beta_{Xjk} + \varepsilon_{Xjk} \nonumber \\
	\hat{\beta}_{Yj} &= \alpha_{j} + \sum_{k=1}^{K}\beta_{Xjk} \theta_{k} + \varepsilon_{Yj} ,
\end{align*}
where $\textrm{var} \left( \varepsilon_{Xjk} \right) = \sigma_{Xjk}^{2}$ and $\textrm{var} \left( \varepsilon_{Yj} \right) = \sigma_{Yj}^{2}$. In two sample Mendelian randomization the genetic variant-risk factor and genetic variant-outcome associations are estimated in separate samples and so $\varepsilon_{Xjk}$ and $\varepsilon_{Yj}$ are independent for all $j$. If $\hat{\beta}_{Xjk}$ and $\hat{\beta}_{Xjl}$ are obtained from separate samples, then $\varepsilon_{Xjk}$ and $\varepsilon_{Xjl}$ are independent. Otherwise, the correlation between $\varepsilon_{Xjk}$ and $\varepsilon_{Xjl}$ depends on the correlation between the $k$\textsuperscript{th} and $l$\textsuperscript{th}  risk factors. Finally, although in practice they are estimated from data, it commonly assumed that $\sigma_{Xjk}^2$ and $\sigma_{Yj}^2$ are known without error for all $j,k$.

Although the model assumes that the risk factors and outcome are continuous, categorical traits are possible, and in fact common in practice. In this case, the relevant genetic variant-trait associations are estimated by logistic regression (or ordinal logistic regression, for ordinal variables with more than two categories) and represent the change in log odds ratio of the trait per extra effect allele in the genetic variant.

\subsection{Genetic variant orientation}
Each genetic variant can be coded in two ways, depending on which allele is chosen as the effect allele. The choice of effect allele is arbitrary, but will change the sign of the genetic variant-trait associations. Some Mendelian randomization methods may give different results depending on the orientation of the genetic variants. For example, in the single risk factor case, the inverse-variance weighted method \citep{Burgess2013IVW} is not affected by genetic variant orientation, but methods which model pleiotropic effects, such as the MR-Egger method \citep{Bowden2015egger} and the lasso-based approach of \citet{Rees2019}, are.

In univariable Mendelian randomization, it is conventional to orientate the genetic variants such that an additional copy of the effect allele has a positive association with the risk factor. In the multiple risk factor case, however, this may be done in multiple ways, since forcing positive associations with respect to one risk factor may change the sign of the associations with respect to the others. \citet{Rees2017MVEgger} suggest orientating the genetic variants such that each one has a positive association with the primary risk factor of interest. If there is no single primary risk factor of interest, or as an additional sensitivity analysis, the impact of the orientation may be assessed by repeating the analysis multiple times, re-orientating the genetic variants with respect to each risk factor.

\subsection{The InSIDE assumption}
An assumption that is often required for pleiotropy robust Mendelian randomization is the InSIDE assumption (instrument strength independent of direct effects) \citep{Bowden2015egger}. There are two forms of the InSIDE assumption: a `population' level version is that, for all $j$, $\alpha_{j}$ is independent of each of $\beta_{Xj1}, \ldots, \beta_{XjK}$; a `finite sample' version is that, for each $j$, the correlation between the sample estimates of $\alpha_{j}$ and each of $\beta_{Xj1}, \ldots, \beta_{XjK}$, for the given set of data, is equal to zero. The latter will rarely be true in practice, since there will typically be residual correlation due to random variation. If the former is true, then this sample correlation will tend to $0$ as the number of instruments increases.

\section{Methods} \label{se:methods}
We proceed to recall existing methods for multivariable Mendelian randomization in Sections \ref{se:ivw} and \ref{se:existingmethods} before introducing new approaches in Sections \ref{se:robust}--\ref{se:reg}.

\subsection{The inverse-variance weighted method} \label{se:ivw}
The multivariable inverse-variance weighted (MVMR-IVW) method \citep{BThompson2015mv, BDudbridgeThompson2015mv} fits the multiple linear regression model
\begin{equation}
	\hat{\beta}_{Yj} = \sum_{k=1}^{K} \hat{\beta}_{Xjk} \theta_{k} + \varepsilon_{j}, \label{eq:mvivw}
\end{equation}
$j = 1, \ldots, p$, where $\varepsilon_{j}$ is normally distributed with mean zero and variance $\sigma_{Yj}^2$. The estimator is obtained using weighted least squares estimation and is thus given by
\begin{equation}
	\argmin_{\theta_{1}, \ldots, \theta_{K}} \sum_{j=1}^{p} \frac{1}{\sigma_{Yj}^2} \left( \hat{\beta}_{Yj} - \sum_{k=1}^{K} \hat{\beta}_{Xjk} \theta_{k} \right)^{2}. \label{eq:ivwls}
\end{equation}
If all genetic variants are valid instruments, $\hat{\theta}_{IVW}$ is a consistent estimator of $\theta$. If not all genetic variants are valid instruments, the estimator remains consistent if pleiotropy is balanced and InSIDE is met. Thus, it is sensitive to outliers and directional pleiotropy.

\subsection{Existing pleiotropy robust methods} \label{se:existingmethods}
If some of the genetic variants are invalid and pleiotropy is directional, the causal effect can still be consistently estimated using the MVMR-Egger method \citep{Rees2017MVEgger}. This method fits an intercept term in (\ref{eq:mvivw}) to account for pleiotropy. That is, we obtain the estimator from
\[
	\argmin_{\theta_{0}, \theta_{1}, \ldots, \theta_{K}} \sum_{j=1}^{p} \frac{1}{\sigma_{Yj}^2} \left( \hat{\beta}_{Yj} - \theta_{0} - \sum_{k=1}^{K} \hat{\beta}_{Xjk} \theta_{k} \right)^{2}. \label{eq:eggerls}
\]
Although the MVMR-Egger estimator is robust to invalidity of instruments, even when all instruments are invalid, it relies on the InSIDE assumption for consistent estimation. Furthermore, it results in lower precision. A final drawback is that it may produce different results depending on the orientation of the genetic variants.

The MR-PRESSO method \citep{Verbanck2018} has been proposed to handle the case where pleiotropy is balanced but there are outliers. Broadly speaking, the method performs a test based on a heterogeneity measure to identify outliers, which are then removed from the analysis. Although \citet{Verbanck2018} describe the method for the single risk factor case, the authors have also produced a multivariable version, which is a straightforward extension. Specifically, the method computes inverse-variance weighted estimates by leaving out one genetic variant at a time. Letting $\hat{\theta}_{1,-j}, \ldots, \hat{\theta}_{K,-j}$ be the estimates obtained after leaving out the $j$\textsuperscript{th} genetic variant, it then computes the following quantity, termed the global observed residual sum of squares:
\[
	\textrm{RSS}_{\textrm{obs}} = \sum_{j=1}^{p} \frac{1}{\sigma_{Yj}^{2}}  \left( \hat{\beta}_{Yj} - \sum_{k=1}^{K} \hat{\beta}_{Xjk} \hat{\theta}_{k, -j} \right)^{2}.
\]
This is compared with an expected residual sum of squares, which is computed multiple ($M$) times:
\[
	\textrm{RSS}_{\textrm{exp}}^{m} = \sum_{j=1}^{p} \frac{1}{\sigma_{Yj}^{2}} \left( \hat{\beta}_{Yj}^{(m)} - \sum_{k=1}^{K} \hat{\beta}_{Xjk}^{(m)} \hat{\theta}_{k, -j} \right)^{2} ,
\]
where $\hat{\beta}_{Xjk}^{(m)}$, $j=1, \ldots, p$, $k=1, \ldots, K$, are drawn from the normal distribution with mean $\hat{\beta}_{Xjk}$ and variance $\sigma_{Xjk}^{2}$, $\hat{\beta}_{Yj}^{(m)}$, $j=1, \ldots, p$, are drawn the normal distribution with mean $\sum_{k=1}^{K} \hat{\beta}_{Xjk} \hat{\theta}_{k, -j}$ and variance $\sigma_{Yj}^{2}$, and $m=1, \ldots, M$. Finally, for each $j$, an empirical p-value is computed as
\[
	\frac{1}{M} \sum_{m=1}^{M} \textbf{1}_{> \textrm{RSS}_{\textrm{obs}}} \left( \textrm{RSS}_{\textrm{exp}}^{m} \right),
\]
where $\textbf{1}_{A} \left( x \right)$ is the indicator function. If the $j$\textsuperscript{th} empirical p-value, multiplied by the number of variants (in order to apply a Bonferroni correction), is greater than the chosen significance level (for example, $0.05$), then the respective genetic variant is identified as an outlier. If there are no outliers identified, the estimate obtained is the same as MVMR-IVW. If true outliers are identified and removed, it is expected to reduce the bias and be more efficient than MVMR-IVW. However, the method is not expected to perform well when there is directional pleiotropy, or there is a large number of invalid instruments.

\subsection{Robust regression} \label{se:robust}

A natural extension to MVMR-IVW is to use robust regression methods, for example MM-estimation. These methods provide robustness to observations which ``contaminate'' the data, such as outliers and influential observations (that is, those for which a small change in observed value results in a large change in parameter estimate). A method for performing robust regression in univariable Mendelian randomization is described in \citet{Rees2019}, which uses MM-estimation along with Tukey's bisquare objective function. It is straightforward to extend this approach to the multivariable model: MM-estimation as described by \citet{KollerStahl2011} is done in a multivariable setting, and it can be implemented using existing software.

This method of robust regression provides robustness to outliers by effectively capping residuals of a certain magnitude. The approach is thus expected to be robust to pleiotropy when there are a relatively small number of invalid instruments. In this case it should be unbiased and more efficient than MVMR-IVW. However, it may not perform well if there are a relatively large number of invalid instruments.

\subsection{Median based estimation} \label{se:median}
An alternative approach to robust regression is to use least absolute deviations regression. That is, we estimate $\theta$ by
\begin{equation}
	\argmin_{\theta_{1}, \ldots, \theta_{K}} \sum_{j=1}^{p} \frac{1}{\sigma_{Yj}^2} \left| \hat{\beta}_{Yj} - \sum_{k=1}^{K} \hat{\beta}_{Xjk} \theta_{k} \right|. \label{eq:lad}
\end{equation}
Least absolute deviations regression is a special case of quantile regression which estimates the $50$\textsuperscript{th} percentile. Thus, (\ref{eq:lad}) is easily computed using techniques developed for quantile regression \citep{Koenker2005}. Since $\hat{\beta}_{Yj}$ and $\hat{\beta}_{Xjk}$ are continuous, (\ref{eq:lad}) has a unique solution with probability one.

Similar to robust regression, least absolute deviations regression is less affected by outliers than least squares regression. It is not robust to influential observations, as robust regression is. However, it may be expected to perform better when the distribution of the $\beta_{Yj}$'s are not symmetric. That is, it also provides robustness to directional pleiotropy. When $K = 1$, the estimator obtained using least absolute deviations regression is equivalent to the weighted median estimator for univariable Mendelian randomization proposed by \citet{Bowden2016median} with weights given by $\lvert \hat{\beta}_{Xj1} \rvert / \sigma_{Yj}^{2}$ (note that, strictly speaking, it is equivalent to the weighted empirical distribution method described in the supplementary material to that paper). The least absolute deviations regression approach can thus be thought of as a natural extension of median-based methods to the multivariable setting. We therefore refer to the method as MVMR-Median.

A disadvantage of least absolute deviations regression is that we lose the asymptotic theory of least squares estimation which leads to easy to compute and accurate standard errors for use, for example, in inference. Confidence intervals are typically produced using a rank inversion technique, or via resampling methods (see, for example, \cite{Tarr2012}). Here we take advantage of the fact that we know the distribution of the genetic variant-trait associations, and implement a parametric bootstrap procedure, as follows. For each genetic variant, a genetic variant-outcome association is drawn from the normal distribution with mean $\hat{\beta}_{Yj}$ and variance $\sigma_{Yj}^{2}$, and genetic variant-risk factor associations are drawn from the multivariate normal distribution with mean $\left( \hat{\beta}_{Xj1}, \ldots, \hat{\beta}_{XjK} \right) '$ and covariance matrix $\textrm{diag} \left( \sigma_{Xj1}^2, \ldots, \sigma_{XjK}^2 \right)$. The estimated standard error is the standard deviation of the estimates computed from multiple replications of this sampling. This approach does not take in to account correlation between the risk factors, however the simulation results presented in Section \ref{se:sims} and the supplementary material show it still performs well in the correlated risk factor case.

\subsection{Regularization methods} \label{se:reg}
Under the assumption that some of the $\alpha_{j}$'s are zero and some are not, regularization methods for univariable Mendelian randomization have been proposed which include an intercept term for each genetic variant in the least squares equations (\ref{eq:ivwls}) and then apply lasso-type penalization to these terms. The penalization tends to shrink the intercept terms corresponding to the valid instruments toward zero. It thus accounts for the pleiotropy caused by invalid instruments, without the loss of power and need for the InSIDE assumption of Egger regression. The approach was first proposed by \citet{Kang2016} in the individual level setting, and followed up by \citet{Windmeijer2019}. \citet{Rees2019} developed a regularization approach using summary level data.

In the multivariable setting we propose using
\begin{equation}
	\argmin_{\theta_{01}, \ldots, \theta_{0p}, \theta_{1}, \ldots, \theta_{K}} \sum_{j=1}^{p} \frac{1}{\sigma_{Yj}^2} \left( \hat{\beta}_{Yj} - \theta_{0j} - \sum_{k=1}^{K} \hat{\beta}_{Xjk} \theta_{k}\right)^{2} + \lambda \sum_{j=1}^{p}\left| \theta_{0j} \right|, \label{eq:mrlasso}
\end{equation}
for some tuning parameter $\lambda > 0$. This is not a standard lasso problem, since not all regression parameters are being penalized. However, the parameter estimates can be easily computed using the algorithm given in Section S.1 of the supplementary material, which uses only standard regression and lasso procedures. The tuning parameter controls the level of sparsity. The larger the value, the fewer genetic variants will be identified as invalid, and the estimate will approach the MVMR-IVW estimate. A data driven approach to choosing the tuning parameter is to use the heterogeneity stopping rule described by \citet{Rees2019}.

The lasso penalty will shrink some $\theta_{0j}$'s exactly to zero, thus identifying the corresponding genetic variants as being valid instruments. A post-lasso estimator takes the genetic variants identified as valid and fits a standard MVMR-IVW model using only these variants. Post-lasso estimators have been advocated by, for example, \citet{Efron2004} and \citet{Belloni2012}, in order to avoid bias caused by the shrinkage of parameter estimates. The lasso algorithm is thus effectively used as a model selection technique.

A limitation of regularization techniques generally is the inability to compute accurate standard errors. We can compute standard errors for the post-lasso estimator using a random effects model \citep{BDudbridgeThompson2015mv} in the post-selection regression. However, this ignores the uncertainty associated with the model selection event. As a result, the standard errors are likely to be too small, and the type I error rate inflated. We examine the effect of this in the simulation study presented in Section \ref{se:sims}. A way around this is to use a three sample approach: here, a set of genetic variant-trait associations is used for performing the MVMR-Lasso procedure which are taken from a sample (or samples) which are independent of those from which the genetic variant-trait associations used for the post-lasso estimator are taken. In this way, the model selection and estimation procedures are independent and the correct type I error rate will be retained \citep{Zhao2019, GB2019}. Although this restricts the potential scope for analyses, since multiple independent samples of genetic variant associations with the traits of interest are required, there are still a number of risk factor-outcome combinations which can be studied given the wide variety of GWAS results which are publicly available. Another promising development in the univariable setting is the use of a selective inference approach, which aims to derive a conditional distribution of the estimator given the model selection event \citep{Bi2019}.

One final point to note is that the solution to (\ref{eq:mrlasso}) may be different depending on the orientation of the genetic variants. Following the convention used when performing MVMR-Egger, we propose orientating the genetic variants such that the genetic variant associations with the primary risk factor of interest are all positive.

\section{Simulations} \label{se:sims}
We conducted a simulation study to compare the performance of the methods described in the previous section under scenarios with different amounts and types of pleiotropy. We simulated from model (\ref{eq:X})--(\ref{eq:Y}) with the intercepts set to zero, $p=100$ genetic variants, $K=4$ risk factors, $n=20\,000$, $\gamma_{Xj} = 1 / K$, $\gamma_{Y} = 1$, $\beta_{Xjk} \sim \textrm{Uniform} \left( 0, 0.1 \right)$, $G_{ij} \sim \textrm{Binomial} \left( 2, 0.3 \right)$,
\[
U_{i} = \sum_{j=1}^{p} \delta_{j} G_{ij} + w_{i}
\]
and $v_{Xi1}, \ldots, v_{XiK}, v_{Yi}, w_{i} \sim N \left( 0, 1 \right)$, independently. These parameter values give $R^{2}$ statistics (that is, the proportion of the variance in each risk factor explained by the genetic variants) of approximately $12\%$. Two sets of values for the causal effects were considered: in the first, $\theta_{1}=0.2, \theta_{2}=0.1, \theta_{3}=0.3, \theta_{4}=0.4$; in the second,  $\theta_{1}=0, \theta_{2}=-0.1, \theta_{3}=0.1, \theta_{4}=0.2$. Three scenarios were considered with different patterns of pleiotropy. For each scenario either $10\%$, $30\%$ or $50\%$ of genetic variants were invalid.
\begin{enumerate}
	\item Balanced pleiotropy and InSIDE assumption met: All $\delta_{j}$'s were set to zero, the proportion of $\alpha_{j}$'s set to zero was either $0.9$, $0.7$ or $0.5$ and non-zero $\alpha_{j}$'s were generated from the $N \left( 0, 0.2^{2} \right)$ distribution.
	\item Directional pleiotropy and InSIDE assumption met: All $\delta_{j}$'s were set to zero, the proportion of $\alpha_{j}$'s set to zero was either $0.9$, $0.7$ or $0.5$ and non-zero $\alpha_{j}$'s were generated from the $N \left( 0.1, 0.2^{2} \right)$ distribution.
	\item Directional pleiotropy and InSIDE assumption violated: All $\alpha_{j}$'s were set to zero, the proportion of $\delta_{j}$'s set to zero was either $0.9$, $0.7$ or $0.5$ and non-zero $\delta_{j}$'s were generated from the $\textrm{Uniform} \left( 0, 0.1 \right)$ distribution (for detail on how the $\delta_{j}$ parameters cause the InSIDE assumption to be violated see, for example, \cite{Bowden20172sample}).
\end{enumerate}

\begin{figure}
	\centering
	\includegraphics{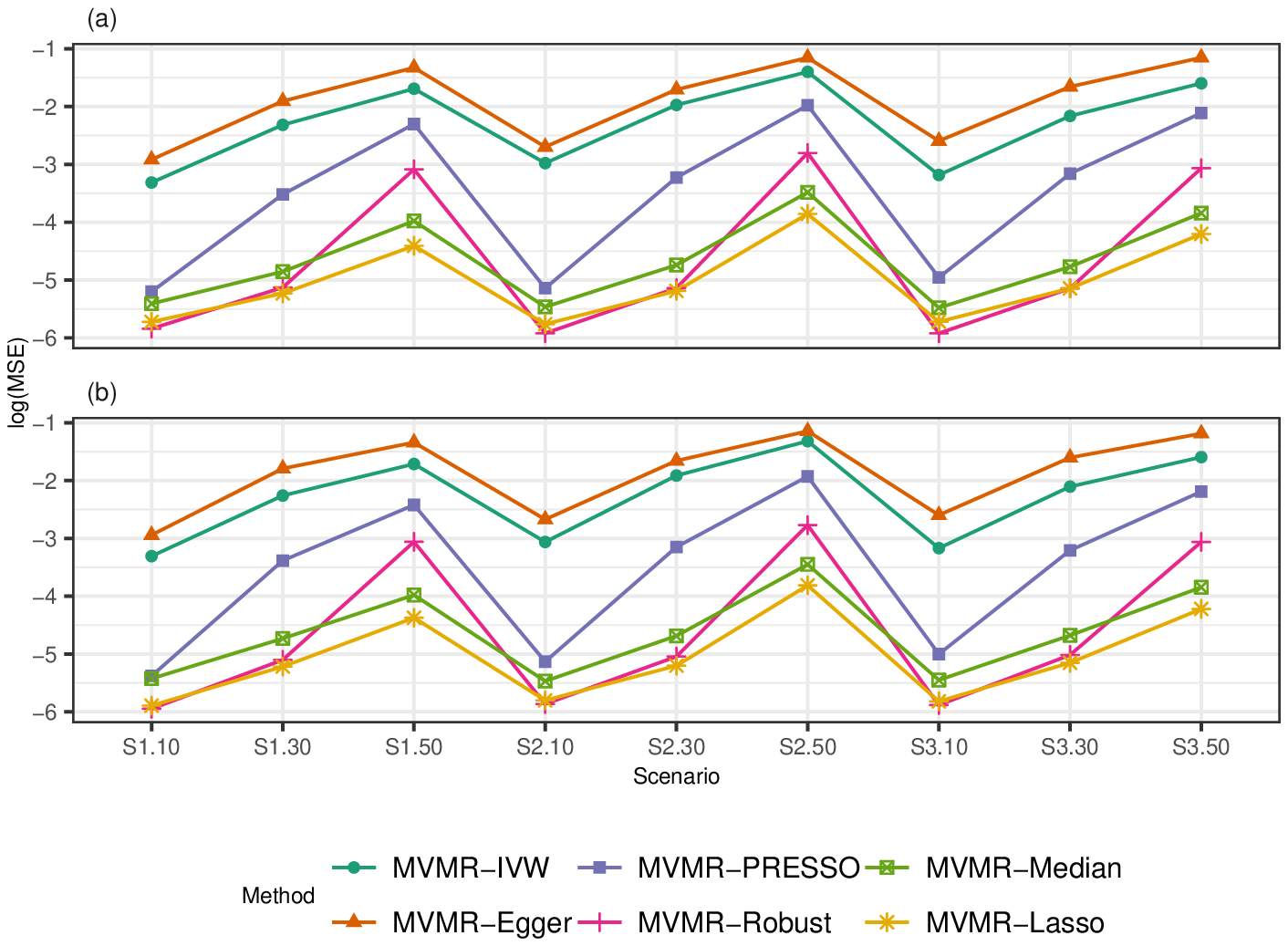}
	\caption{Logarithm of the mean squared errors for each scenario (S1, S2 and S3) and proportion of invalid genetic variants (10, 30 or 50\%), where (a) $\theta_{1} = 0.2$ and (b) $\theta_{1} = 0$.}
	\label{fg:mse}
\end{figure}

For each scenario, level of pleiotropy, and set of $\theta_{k}$ values, the simulations were replicated $1\,000$ times. For each replication, the genetic variant-trait association estimates and their standard errors were computed from the individual level data using simple linear regression with an intercept. The causal effects were then estimated using the methods described in Section \ref{se:methods}. The parameter of interest which we report on was the causal effect of the first risk factor on the outcome (that is, for the first set of $\theta_{k}$ values, there is a true causal effect, and for the second set of values there is no causal effect). The mean, standard deviation of estimates, mean standard error and power / type I error rate, at the $0.05$ significance level, are shown in Tables \ref{tb:simsHA}--\ref{tb:simsH0}. The log of the mean squared errors across all scenarios are shown in Figure \ref{fg:mse}. Note that here the MVMR-Lasso method refers to the two sample post-lasso estimator (that is, with the estimate computed from the same samples that the instruments were selected in).

All methods performed well in terms of bias when there was balanced pleiotropy. The MVMR-IVW and MVMR-Egger methods were biased when pleiotropy was directional, increasing as the proportion of pleiotropy increased. These methods were also less precise than all other methods, with the largest standard deviations of estimates, and were very low powered. In theory, MVMR-Egger should be robust to directional pleiotropy when InSIDE is met. However, there was a fair amount of bias in these scenarios. An explanation for this is that the bias is due to weak instruments, which this method is particularly susceptible to. This was further examined in the supplementary simulations (see the discussion at the end of this section).

MVMR-Robust outperformed MVMR-PRESSO in all scenarios with lower bias, more precision and correct type I errors rates. MVMR-PRESSO had low bias at the lower level of pleiotropy, but did not perform well with moderate or high amounts. MVMR-Lasso was generally the most precise estimate: it had similar mean squared error to MVMR-Robust at $10\%$ pleiotropy, but retained its performance in this regard at the higher levels of pleiotropy also. Similarly, it had comparable power to MVMR-Robust at $10\%$ pleiotropy, but did much better as the proportion of pleiotropy increased. As expected, MVMR-Lasso had inflated type I error rates. MVMR-Median had comparable bias to MVMR-Lasso across all levels of pleiotropy. It was less precise and lower powered than MVMR-Lasso, but had type I error rates closer to the significance level. In terms of mean squared error, MVMR-Median was bettered uniformly across all scenarios only by MVMR-Lasso. As a further analysis, in Figure S1 in the supplementary material, we compare the mean squared errors with those from the estimates of the causal effect of the fourth risk factor, that is, where the true causal effect is $\theta_{4} = 0.4$. The average proportion of variation in the outcome explained by the fourth risk factor is approximately $19\%$, compared with approximately $11\%$ for the first risk factor. However, with the exception of MVMR-Egger, the results are almost identical across all scenarios.

The simulations were repeated for the cases where there were fewer instruments ($p = 20$, with the distribution of the $\beta_{Xjk}$ parameters adjusted to retain similar $R^{2}$ values), where the risk factors were correlated by setting $\textrm{cor} \left( v_{Xik}, v_{Xil} \right) = 0.5$ for all $k \neq l$, and where the genetic variant-trait associations were all estimated from the same sample (one sample Mendelian randomization). The results are shown in Tables \ref{tb:simsHAp20}--\ref{tb:simsH0S1} and Figures \ref{fg:msep20}--\ref{fg:mseS1}. In each case, the results followed a similar pattern as before in terms of comparative performance of the different methods. In the fewer instruments case, there was slightly higher mean squared error across the board, which would be expected with fewer instruments, but the differences were not great. The results also support the assertion that the bias from MVMR-Egger shown in Tables \ref{tb:simsHA}--\ref{tb:simsH0} is due to weak instruments, since in this case the genetic variant-risk factor associations were adjusted to control the $R^{2}$ values, but the average F statistics were higher. The results from the correlated risk factor case suggested that the parametric bootstrap procedure (which effectively ignores risk factor correlation) still performs well.

\begin{table}[p]
	\centering
	\caption{Mean and standard deviation (SD) of estimates, mean standard error of estimates (SE) and power when $\theta_{1}=0.2$.}
	\label{tb:simsHA}
	\scalebox{0.8}{
		\begin{tabular}{l c c c c c c c c c c c c c c}
			\hline
			& \multicolumn{4}{c}{$10\%$ invalid} & & \multicolumn{4}{c}{$30\%$ invalid} & & \multicolumn{4}{c}{$50\%$ invalid} \\ \cline{2-5} \cline{7-10} \cline{12-15}
			Method & Mean & SD & SE & Power & & Mean & SD & SE &  Power & &Mean & SD & SE & Power \\ \hline
			& \multicolumn{14}{c}{Scenario 1: Balanced pleiotropy, InSIDE met} \\ \hline
			MVMR-IVW&0.201&0.191&0.185&0.222&&0.202&0.314&0.321&0.094&&0.198&0.430&0.415&0.089
\\
			MVMR-Egger&0.207&0.233&0.226&0.190&&0.219&0.385&0.391&0.084&&0.207&0.515&0.502&0.074
\\
			MVMR-PRESSO&0.206&0.074&0.067&0.824&&0.201&0.172&0.141&0.373&&0.186&0.316&0.221&0.243
\\
			MVMR-Robust&0.205&0.054&0.055&0.958&&0.203&0.077&0.078&0.735&&0.186&0.214&0.235&0.128
\\
			MVMR-Median&0.206&0.067&0.082&0.744&&0.202&0.088&0.100&0.528&&0.197&0.137&0.127&0.367
\\
			MVMR-Lasso&0.205&0.057&0.058&0.934&&0.204&0.073&0.068&0.830&&0.199&0.110&0.083&0.652\\ \hline
			& \multicolumn{14}{c}{Scenario 2: Directional pleiotropy, InSIDE met} \\ \hline
			MVMR-IVW&0.256&0.219&0.206&0.272&&0.332&0.349&0.350&0.171&&0.416&0.448&0.442&0.159
\\
			MVMR-Egger&0.240&0.257&0.251&0.192&&0.267&0.421&0.424&0.105&&0.300&0.553&0.536&0.086
\\
			MVMR-PRESSO&0.208&0.076&0.070&0.837&&0.239&0.196&0.157&0.390&&0.334&0.347&0.253&0.313
\\
			MVMR-Robust&0.203&0.052&0.055&0.963&&0.201&0.077&0.080&0.733&&0.256&0.240&0.264&0.144
\\
			MVMR-Median&0.207&0.065&0.084&0.733&&0.214&0.093&0.105&0.543&&0.247&0.169&0.142&0.425
\\
			MVMR-Lasso&0.204&0.056&0.059&0.939&&0.205&0.075&0.071&0.809&&0.225&0.143&0.089&0.661\\ \hline
			& \multicolumn{14}{c}{Scenario 3: Directional pleiotropy, InSIDE violated} \\ \hline
			MVMR-IVW&0.238&0.200&0.192&0.276&&0.283&0.329&0.329&0.132&&0.313&0.436&0.419&0.121
\\
			MVMR-Egger&0.276&0.263&0.231&0.266&&0.352&0.410&0.390&0.155&&0.385&0.531&0.491&0.145
\\
			MVMR-PRESSO&0.210&0.083&0.074&0.800&&0.245&0.201&0.164&0.387&&0.285&0.338&0.258&0.270
\\
			MVMR-Robust&0.203&0.052&0.056&0.963&&0.205&0.076&0.079&0.748&&0.230&0.214&0.239&0.153
\\
			MVMR-Median&0.206&0.064&0.083&0.763&&0.214&0.091&0.102&0.570&&0.225&0.144&0.133&0.417
\\
			MVMR-Lasso&0.204&0.057&0.059&0.947&&0.206&0.076&0.069&0.818&&0.213&0.122&0.085&0.657\\ \hline
	\end{tabular}}
\end{table}

\begin{table}[p]
	\centering
	\caption{Mean and standard deviation (SD) of estimates, mean standard error of estimates (SE) and type I error rate when $\theta_{1}=0$.}
	\label{tb:simsH0}
	\scalebox{0.8}{
		\begin{tabular}{l c c c c c c c c c c c c c c}
			\hline
			& \multicolumn{4}{c}{$10\%$ invalid} & & \multicolumn{4}{c}{$30\%$ invalid} & & \multicolumn{4}{c}{$50\%$ invalid} \\ \cline{2-5} \cline{7-10} \cline{12-15}
			Method & Mean & SD & SE & Type I & & Mean & SD & SE & Type I & &Mean & SD & SE & Type I \\ \hline
			& \multicolumn{14}{c}{Scenario 1: Balanced pleiotropy, InSIDE met} \\ \hline
			MVMR-IVW&0.003&0.191&0.187&0.054&&0.014&0.323&0.321&0.045&&0.016&0.424&0.412&0.053
\\
			MVMR-Egger&0.015&0.229&0.228&0.047&&0.022&0.408&0.391&0.054&&0.017&0.511&0.501&0.055
\\
			MVMR-PRESSO&0.003&0.068&0.064&0.053&&0.003&0.184&0.136&0.098&&-0.003&0.298&0.214&0.100\\
			MVMR-Robust&0.002&0.051&0.051&0.053&&0.000&0.078&0.078&0.049&&-0.001&0.217&0.235&0.028
\\
			MVMR-Median&0.003&0.066&0.071&0.038&&0.006&0.094&0.088&0.054&&0.001&0.137&0.114&0.093
\\
			MVMR-Lasso&0.002&0.053&0.051&0.064&&0.002&0.074&0.061&0.096&&0.002&0.112&0.075&0.171\\ \hline
			& \multicolumn{14}{c}{Scenario 2: Directional pleiotropy, InSIDE met} \\ \hline
			MVMR-IVW&0.044&0.212&0.206&0.065&&0.123&0.364&0.350&0.070&&0.253&0.452&0.441&0.093
\\
			MVMR-Egger&0.026&0.262&0.250&0.058&&0.051&0.434&0.425&0.059&&0.123&0.551&0.535&0.068
\\
			MVMR-PRESSO&0.007&0.077&0.067&0.064&&0.042&0.203&0.157&0.088&&0.152&0.350&0.245&0.173
\\
			MVMR-Robust&0.003&0.053&0.052&0.072&&0.005&0.080&0.079&0.050&&0.070&0.240&0.266&0.023
\\
			MVMR-Median&0.006&0.065&0.072&0.026&&0.019&0.094&0.092&0.060&&0.056&0.169&0.129&0.126
\\
			MVMR-Lasso&0.003&0.055&0.052&0.063&&0.006&0.074&0.063&0.093&&0.033&0.145&0.082&0.233\\ \hline
			& \multicolumn{14}{c}{Scenario 3: Directional pleiotropy, InSIDE violated} \\ \hline
			MVMR-IVW&0.043&0.201&0.191&0.057&&0.094&0.336&0.330&0.079&&0.131&0.431&0.419&0.077
\\
			MVMR-Egger&0.080&0.261&0.230&0.082&&0.164&0.418&0.391&0.099&&0.192&0.518&0.491&0.088
\\
			MVMR-PRESSO&0.013&0.081&0.070&0.068&&0.037&0.198&0.158&0.081&&0.081&0.325&0.249&0.115
\\
			MVMR-Robust&0.008&0.052&0.051&0.057&&0.013&0.080&0.079&0.051&&0.028&0.215&0.241&0.024
\\
			MVMR-Median&0.011&0.065&0.072&0.030&&0.019&0.095&0.090&0.066&&0.025&0.144&0.120&0.104
\\
			MVMR-Lasso&0.008&0.054&0.051&0.068&&0.013&0.075&0.062&0.099&&0.014&0.120&0.077&0.183\\ \hline
	\end{tabular}}
\end{table}

\section{Applied example: the causal effect of intelligence, education and household income on Alzheimer's disease}
In this section we consider an applied example looking at the causal effects of intelligence, years of education and household income on Alzheimer's disease. The effects of intelligence and years of education on health outcomes have been studied by \citet{Davies2019} and \citet{Anderson2020}. A multivariable approach is important in this case since intelligence and years of education are highly correlated. \citet{Anderson2020} used univariable Mendelian randomization with intelligence and years of education, separately, as risk factors, and Alzheimer's disease as the outcome. The results suggest that both risk factors have a protective effect on Alzheimer's disease. However, when both risk factors are included in a multivariable model, using MVMR-IVW, the effect of years of education, independent of intelligence, shifts toward the null. The implication is that years of education only has a causal effect on the odds ratio of Alzheimer's disease via its effect on intelligence. Here, we reconsider this example and include household income as an extra risk factor.

Genetic variant associations with intelligence and years of education are taken from the GWAS of \citet{Hill2019} and \citet{Okbay2016}, respectively. By clumping the combined list of genetic variants which associate with each risk factor at the genome wide significance level, \citet{Davies2019} arrived at a list of 219 independent genetic variants to be used as instruments in multivariable Mendelian randomization analyses. We obtained the associations between these genetic variants and household income from the UK Biobank (sourced from \url{http://www.nealelab.is/uk-biobank/}). Note that household income is an ordinal categorical variable, and so the genetic variant associations represent the increase in log odds of being in a higher income category per extra effect allele. Genetic variant associations with Alzheimer's disease were obtained from the GWAS of \citet{Lambert2013}. In total, 213 of the genetic variants used by \citet{Davies2019} were available in both of the household income and Alzheimer's disease datasets, and we used these as instruments in our analysis. Note that the genetic variant associations with both intelligence and years of education were all in the same direction, and so they were orientated in our analysis to be all positive with respect to these traits.

\begin{figure}
	\centering
	\includegraphics{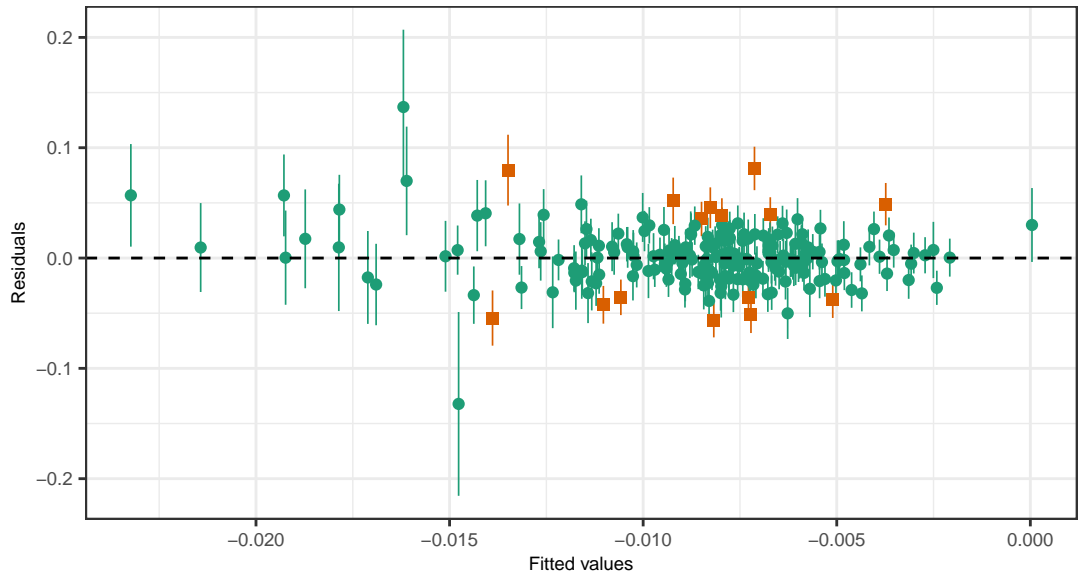}
	\caption{Residuals vs fitted values from the MVMR-IVW model. The vertical error bars indicate $\pm \sigma_{Yj}$ for each genetic variant. The orange box shaped points indicate the genetic variants identified as pleiotropic by the MVMR-Lasso method.}
	\label{fg:resfitplot}
\end{figure}

\begin{figure}
	\centering
	\includegraphics{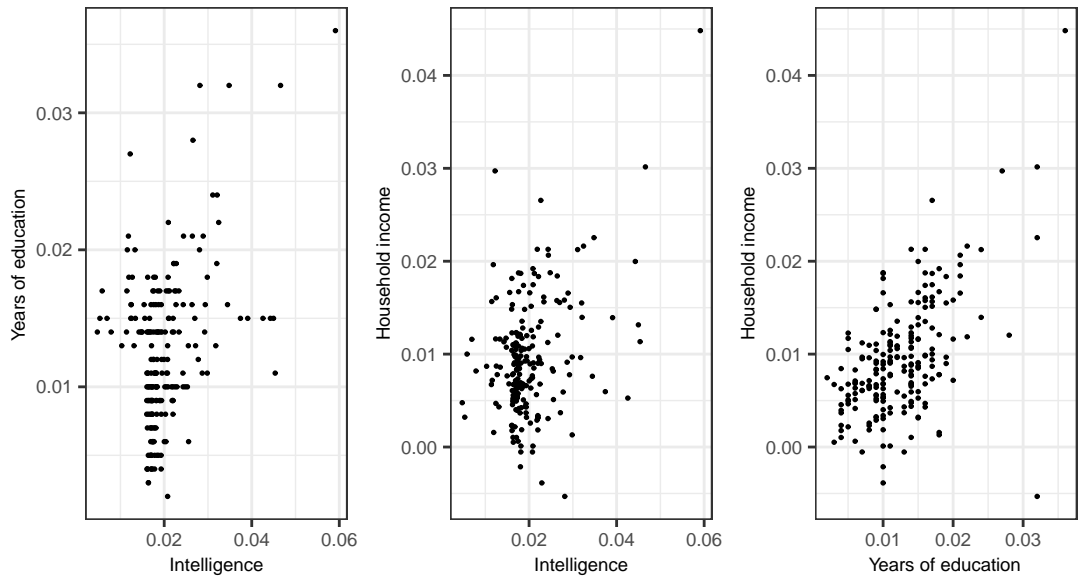}
	\caption{Scatterplots of each pair of genetic variant associations with intelligence, years of education and household income.}
	\label{fg:bxgplots}
\end{figure}

Figure \ref{fg:resfitplot} shows a plot of the residuals vs fitted values after fitting the MVMR-IVW model to the data. The vertical error bars indicate $\pm \sigma_{Yj}$ for each genetic variant. The plot provides a way of visualising heterogeneity in the multivariable setting, similar to the scatterplots of $\hat{\beta}_{Xj}$ against $\hat{\beta}_{Yj}$ commonly used in the univariable case. Although there is little evidence of directional pleiotropy, there may be some outliers. Figure \ref{fg:bxgplots} shows scatterplots of each pair of genetic variant-risk factor associations. There appears to be reasonably strong correlation between the genetic variant associations with years of education and household income, and low to moderate correlation between the other two pairs of associations.

\begin{figure}
	\centering
	\includegraphics{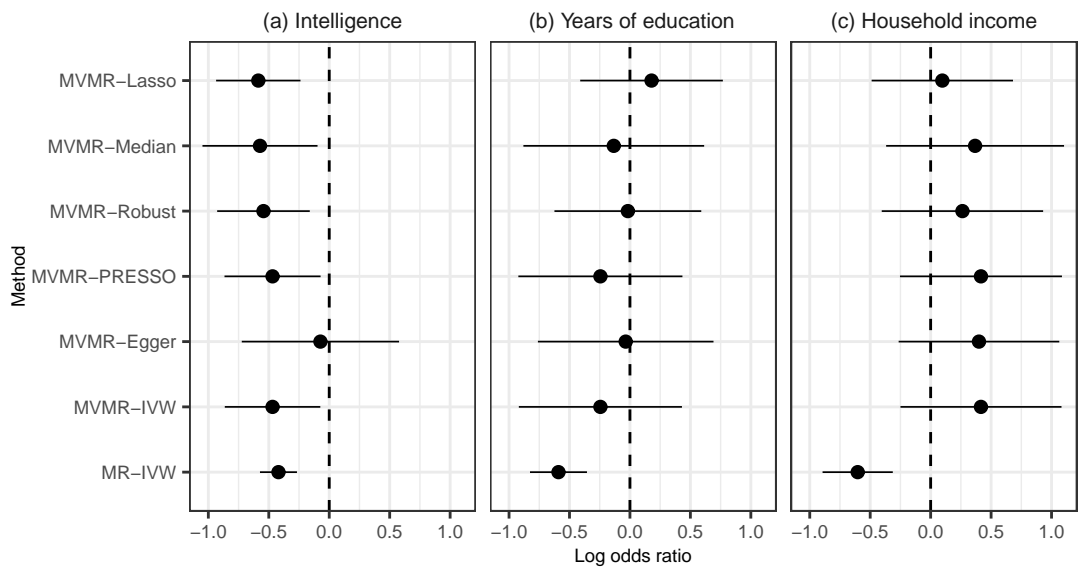}
	\caption{Log causal odds ratio for Alzheimer's disease point estimate and 95\% confidence interval per one standard deviation increase in: (a) intelligence; (b) years of education and; (c) per unit increase in log odds ratio for a higher household income bracket.}
	\label{fg:mrestplots}
\end{figure}

We used the univariable inverse-variance weighted method to estimate the causal effect of each risk factor separately on the odds ratio of Alzheimer's disease. We then applied each of the multivariable methods discussed in Section \ref{se:methods} with all three risk factors included. The results are shown in Figure \ref{fg:mrestplots}(a)--(c).

The univariable analyses suggest that intelligence and years of education both have a protective effect on Alzheimer's disease, in line with the results of \citet{Anderson2020}. The estimated log causal odds ratio of Alzheimer's disease per one standard deviation increase in intelligence is $-4.20$ (95\% CI $-0.57$ to $-0.27$), and per one standard deviation increase in years of education is $-0.59$ (95\% CI $-0.83$ to $-0.36$). The estimated log causal odds ratio of Alzheimer's disease per unit increase in log odds ratio for a higher household income bracket is $-0.60$ (95\% CI $-0.89$ to $-0.31$). Using the MVMR-IVW model, the estimates of the log causal odds ratio from both years of education and household income attenuated to the null, with 95\% confidence intervals overlapping zero. The multivariable model however still suggests a protective effect from intelligence, with an estimated odds ratio of $-0.47$ (95\% CI $-0.86$ to $-0.07$).

The pleiotropy robust multivariable methods gave results which were broadly consistent with the MVMR-IVW results (see Table \ref{tb:mvmr_results}). The MVMR-Egger method suggested a null causal effect of intelligence on Alzheimer's disease, however the point estimate is still in the same direction and all other methods were in line with MVMR-IVW. Note that the MR-PRESSO outlier test did not detect any outliers, but the MVMR-Lasso method identified $15$ genetic variants as pleiotropic, which were removed before computing the post-lasso estimator. These genetic variants are indicated in Figure \ref{fg:resfitplot}. Interestingly, the post-lasso estimate for the causal effect of years of education on Alzheimer's disease was positive, whereas all other estimates of this effect were negative. However, the confidence interval still included the null.

The consistency of the findings give strength to the assertion that intelligence has a causally protective effect on Alzheimer's disease, independent of years of education and household income. There are two potential explanations for the findings relating to years of education and household income. One is that these risk factors affect Alzheimer's disease via their association with intelligence. That is, intelligence is a mediator of the effect of years of education and/or household income. The other is that these risk factors sit on pleiotropic pathways between the genetic variants and Alzheimer's disease, not necessarily passing through intelligence. It is also possible that both explanations are true (that is, that there is both pleiotropic effects and mediation via intelligence).

\begin{table}[p]
	\centering
	\caption{Point estimate, standard error (SE) and confidence interval (CI Lower, CI Upper) of the log odds ratio of Alzheimer's disease due to a unit increase in intelligence, years of education and household income, from univariable Mendelian randomization (MR-IVW) and each multivariable method.}
	\label{tb:mvmr_results}
	\scalebox{1}{
		\begin{tabular}{l l c c c c}
			\hline
			Risk factor & Method & Estimate & SE & CI Lower & CI Upper\\ \hline
			Intelligence & MR-IVW & -0.420 & 0.078 & -0.573 & -0.267 \\
			& MVMR-IVW&-0.469&0.202&-0.864&-0.074
			\\
			&MVMR-Egger&-0.073&0.332&-0.723&0.578
			\\
			&MVMR-PRESSO&-0.469&0.202&-0.866&-0.072\\
			&MVMR-Robust&-0.544&0.195&-0.927&-0.161
			\\
			&MVMR-Median&-0.573&0.241&-1.045&-0.100
			\\
			&MVMR-Lasso&-0.587&0.178&-0.936&-0.238\\ \hline
			Years of education&MR-IVW&-0.591&0.120&-0.827&-0.355
			\\
			&MVMR-IVW&-0.244&0.344&-0.919&0.430
			\\
			&MVMR-Egger&-0.035&0.371&-0.761&0.691
			\\
			&MVMR-PRESSO&-0.244&0.344&-0.923&0.434
			\\
			&MVMR-Robust&-0.017&0.310&-0.624&0.590
			\\
			&MVMR-Median&-0.134&0.384&-0.887&0.620
			\\
			&MVMR-Lasso&0.179&0.301&-0.411&0.769\\ \hline
			Household income&MR-IVW&-0.603&0.148&-0.894&-0.313
			\\
			&MVMR-IVW&0.416&0.340&-0.250&1.082
			\\
			&MVMR-Egger&0.400&0.339&-0.265&1.064
			\\
			&MVMR-PRESSO&0.416&0.340&-0.254&1.086
			\\
			&MVMR-Robust&0.263&0.341&-0.404&0.931
			\\
			&MVMR-Median&0.368&0.381&-0.378&1.114
			\\
			&MVMR-Lasso&0.097&0.298&-0.488&0.681\\ \hline
	\end{tabular}}
\end{table}

\section{Discussion}
In this paper we have presented methods for performing multivariable Mendelian randomization which are robust to pleiotropy. Existing methods either allow for invalidity at the cost of low precision and the InSIDE assumption (MVMR-Egger), or were developed for the case where pleiotropy is balanced and there are a relatively small number of outliers (MVMR-PRESSO). We have considered methods which can handle higher proportions of invalidity and directional pleiotropy.

When there is evidence of relatively few invalid instruments, MVMR-Robust was shown to outperform MVMR-PRESSO in all scenarios considered. MVMR-Lasso, another method which aims to identify and downweight outliers, performed best overall in terms of mean squared error, even when half of the genetic variants were invalid instruments and pleiotropy was directional. Although type I error rates were inflated, this can be mitigated when a three sample approach is possible. MVMR-Median was shown to perform almost as well as MVMR-Lasso in terms of mean squared error, and retained correct type I error rates at higher levels of pleiotropy. As demonstrated in the applied example, a plot of the residuals versus fitted values from the MVMR-IVW method can be used to visualise potential outliers and pleiotropy, and to help determine the most appropriate choice of robust method.

The work has some limitations in the modelling assumptions made, in particular of the linearity and homogeneity (that is, no effect modification) of the effects of the risk factors on the outcome. Furthermore, although we can handle non-continuous traits via the use of logistic regression to produce summary statistics, this may cause bias in the causal effect estimates due to the non-collapsibility of the odds ratio. Nonetheless, violations of these assumptions tend to attenuate causal effect estimates toward the null \citep{Bowden20172sample, BDudbridge2016allele}.

Another limitation is the assumption of no measurement error of the genetic variant-risk factor associations, equivalent to assuming $\sigma_{Xjk} = 0$ for all $j, k$. This is a common assumption in Mendelian randomization analyses, and is justified by the very large sample sizes that these associations are typically estimated in, in contrast to the genetic variant-outcome associations which may be estimated using a relatively small number of cases vs controls. Provided the genetic variants strongly predict each risk factor, conditional on all the other risk factors, this assumption will have little influence on the analysis. Otherwise, the results may be subject to weak instrument bias. In practice, weak instrument bias is mitigated by selecting a set of genetic variants which associate with the risk factors according to some threshold related to, for example, an F-statistic or p-value. \citet{Sanderson2020} provide a test to help to diagnose (conditionally) weak instruments and propose an approach to adjust standard multivariable Mendelian randomization estimators to account for the $\sigma_{Xjk}$'s being non-zero. Although in univariable Mendelian randomization measurement error will bias causal effect estimates toward the null \citep{BurgessThompson2011}, this will not necessarily be the case in the multivariable setting. Assessing the impact of measurement error in the multiple risk factor case, and how to account for this, is an active area of research.

In summary, the methods we have presented provide new ways for performing Mendelian randomization with multiple risk factors which are robust to different forms of pleiotropy. Each has advantages when applied to specific scenarios. Together with MVMR-Egger, these methods provide a suite of sensitivity analyses for multivariable Mendelian randomization.

\section{Software}
R code for performing the methods described in this paper, and for reproducing the simulation results, can be found at \url{https://github.com/aj-grant/robust-mvmr}. Existing R packages that are used to implement the various methods include: MendelianRandomization \citep{mrpackage}; MR-PRESSO \citep{Verbanck2018}; robustbase \citep{robustbase}; quantreg \citep{quantreg}; and glmnet \citep{glmnet}.

\section*{Acknowledgments}
The authors thank Neil Davies for supplying some of the data used in the applied example, and Max Mandl for supplying a correction to the multivariable version of the mr\_presso R function.

\section*{Funding}
Andrew J. Grant and Stephen Burgess are supported by a Sir Henry Dale Fellowship jointly funded by the Wellcome Trust and the Royal Society (grant number 204623/Z/16/Z). This research was funded by the NIHR Cambridge Biomedical Research Centre. The views expressed are those of the authors and not necessarily those of the NHS, the NIHR or the Department of Health and Social Care.

\bibliographystyle{unsrtnat_ag}
\bibliography{robust_mvmr_abb_bib}	

\begin{thebibliography}{38}
\providecommand{\natexlab}[1]{#1}
\providecommand{\url}[1]{\texttt{#1}}
\expandafter\ifx\csname urlstyle\endcsname\relax
  \providecommand{\doi}[1]{doi: #1}\else
  \providecommand{\doi}{doi: \begingroup \urlstyle{rm}\Url}\fi

\bibitem[Davey~Smith and Ebrahim(2003)]{GDS2003}
Davey~Smith, G. and Ebrahim, S.
\newblock {‘Mendelian randomization’: can genetic epidemiology contribute
  to understanding environmental determinants of disease?}
\newblock \emph{Int J Epidemiol}, 32\penalty0 (1):\penalty0 1--22, 2003.

\bibitem[Greenland(2000)]{Greenland2000}
Greenland, S.
\newblock {An introduction to instrumental variables for epidemiologists}.
\newblock \emph{Int J Epidemiol}, 29\penalty0 (4):\penalty0 722--729, 2000.

\bibitem[Burgess et~al.(2013{\natexlab{a}})Burgess, Butterworth, and
  Thompson]{BButterworthThompson2013}
Burgess, S., Butterworth, A., and Thompson, S.~G.
\newblock {M}endelian randomization analysis with multiple genetic variants
  using summarized data.
\newblock \emph{Genet Epidemiol}, 37\penalty0 (7):\penalty0 658--665,
  2013{\natexlab{a}}.

\bibitem[Pierce and Burgess(2013)]{PierceBurgess2013}
Pierce, B.~L. and Burgess, S.
\newblock {Efficient Design for {M}endelian Randomization Studies: Subsample
  and 2-Sample Instrumental Variable Estimators}.
\newblock \emph{Am J Epidemiol}, 178\penalty0 (7):\penalty0 1177--1184, 07
  2013.

\bibitem[Burgess and Thompson(2015)]{BThompson2015mv}
Burgess, S. and Thompson, S.~G.
\newblock Multivariable {M}endelian randomization: the use of pleiotropic
  genetic variants to estimate causal effects.
\newblock \emph{Am J Epidemiol}, 181\penalty0 (4):\penalty0 251--260, 2015.

\bibitem[Sanderson et~al.(2019)Sanderson, Davey~Smith, Windmeijer, and
  Bowden]{Sanderson2019}
Sanderson, E., Davey~Smith, G., Windmeijer, F., and Bowden, J.
\newblock {An examination of multivariable Mendelian randomization in the
  single-sample and two-sample summary data settings}.
\newblock \emph{Int J Epidemiol}, 48\penalty0 (3):\penalty0 713--727, 2019.

\bibitem[Slob and Burgess(2020)]{Slob2020}
Slob, E. A.~W. and Burgess, S.
\newblock A comparison of robust mendelian randomization methods using summary
  data.
\newblock \emph{Genet Epidemiol}, 44\penalty0 (4):\penalty0 313--329, 2020.

\bibitem[Lawlor et~al.(2017)Lawlor, Tilling, and Davey~Smith]{Lawlor2017}
Lawlor, D.~A., Tilling, K., and Davey~Smith, G.
\newblock {Triangulation in aetiological epidemiology}.
\newblock \emph{Int J Epidemiol}, 45\penalty0 (6):\penalty0 1866--1886, 2017.

\bibitem[Bowden et~al.(2017)Bowden, Del Greco~M, Minelli, Davey~Smith,
  et~al.]{Bowden20172sample}
Bowden, J., Del Greco~M, F., Minelli, C., Davey~Smith, G., et~al.
\newblock A framework for the investigation of pleiotropy in two-sample summary
  data {M}endelian randomization.
\newblock \emph{Stat Med}, 36\penalty0 (11):\penalty0 1783--1802, 2017.

\bibitem[Burgess et~al.(2013{\natexlab{b}})Burgess, Butterworth, and
  Thompson]{Burgess2013IVW}
Burgess, S., Butterworth, A., and Thompson, S.~G.
\newblock Mendelian randomization analysis with multiple genetic variants using
  summarized data.
\newblock \emph{Genet Epidemiol}, 37\penalty0 (7):\penalty0 658--665,
  2013{\natexlab{b}}.

\bibitem[Bowden et~al.(2015)Bowden, Davey~Smith, and Burgess]{Bowden2015egger}
Bowden, J., Davey~Smith, G., and Burgess, S.
\newblock {Mendelian randomization with invalid instruments: effect estimation
  and bias detection through Egger regression}.
\newblock \emph{Int J Epidemiol}, 44\penalty0 (2):\penalty0 512--525, 2015.

\bibitem[Rees et~al.(2019)Rees, Wood, Dudbridge, and Burgess]{Rees2019}
Rees, J. M.~B., Wood, A.~M., Dudbridge, F., and Burgess, S.
\newblock Robust methods in {M}endelian randomization via penalization of
  heterogeneous causal estimates.
\newblock \emph{PLoS ONE}, 14\penalty0 (9):\penalty0 1--24, 2019.

\bibitem[Rees et~al.(2017)Rees, Wood, and Burgess]{Rees2017MVEgger}
Rees, J. M.~B., Wood, A.~M., and Burgess, S.
\newblock Extending the {MR}-egger method for multivariable {M}endelian
  randomization to correct for both measured and unmeasured pleiotropy.
\newblock \emph{Stat Med}, 36\penalty0 (29):\penalty0 4705--4718, 2017.

\bibitem[Burgess et~al.(2015)Burgess, Dudbridge, and
  Thompson]{BDudbridgeThompson2015mv}
Burgess, S., Dudbridge, F., and Thompson, S.~G.
\newblock {Re: “Multivariable {M}endelian randomization: the use of
  pleiotropic genetic variants to estimate causal effects”}.
\newblock \emph{Am J Epidemiol}, 181\penalty0 (4):\penalty0 290--291, 2015.

\bibitem[Verbanck et~al.(2018)Verbanck, Chen, Neale, and Do]{Verbanck2018}
Verbanck, M., Chen, C.-Y., Neale, B., and Do, R.
\newblock {Detection of widespread horizontal pleiotropy in causal
  relationships inferred from {M}endelian randomization between complex traits
  and diseases}.
\newblock \emph{Nat. Genet.}, 50\penalty0 (5):\penalty0 693--698, 2018.

\bibitem[Koller and Stahel(2011)]{KollerStahl2011}
Koller, M. and Stahel, W.~A.
\newblock Sharpening {W}ald-type inference in robust regression for small
  samples.
\newblock \emph{Comput Stat Data Anal}, 55\penalty0 (8):\penalty0 2504 -- 2515,
  2011.

\bibitem[Koenker(2005)]{Koenker2005}
Koenker, R.
\newblock \emph{Quantile Regression}.
\newblock Cambridge University Press, Cambridge, 2005.

\bibitem[Bowden et~al.(2016)Bowden, Davey~Smith, Haycock, and
  Burgess]{Bowden2016median}
Bowden, J., Davey~Smith, G., Haycock, P.~C., and Burgess, S.
\newblock Consistent estimation in {M}endelian randomization with some invalid
  instruments using a weighted median estimator.
\newblock \emph{Genet Epidemiol}, 40\penalty0 (4):\penalty0 304--314, 2016.

\bibitem[Tarr(2012)]{Tarr2012}
Tarr, G.
\newblock Small sample performance of quantile regression confidence intervals.
\newblock \emph{J Stat Comput Simul}, 82\penalty0 (1):\penalty0 81--94, 2012.

\bibitem[Kang et~al.(2016)Kang, Zhang, Cai, and Small]{Kang2016}
Kang, H., Zhang, A., Cai, T.~T., and Small, D.~S.
\newblock Instrumental variables estimation with some invalid instruments and
  its application to {M}endelian randomization.
\newblock \emph{J Am Stat Assoc}, 111\penalty0 (513):\penalty0 132--144, 2016.

\bibitem[Windmeijer et~al.(2019)Windmeijer, Farbmacher, Davies, and
  Davey~Smith]{Windmeijer2019}
Windmeijer, F., Farbmacher, H., Davies, N., and Davey~Smith, G.
\newblock On the use of the {L}asso for instrumental variables estimation with
  some invalid instruments.
\newblock \emph{J Am Stat Assoc}, 114\penalty0 (527):\penalty0 1339--1350,
  2019.

\bibitem[Efron et~al.(2004)Efron, Hastie, Johnstone, and Tibshirani]{Efron2004}
Efron, B., Hastie, T., Johnstone, I., and Tibshirani, R.
\newblock Least angle regression.
\newblock \emph{Ann Stat}, 32\penalty0 (2):\penalty0 407--499, 2004.

\bibitem[Belloni et~al.(2012)Belloni, Chen, Chernozhukov, and
  Hansen]{Belloni2012}
Belloni, A., Chen, D., Chernozhukov, V., and Hansen, C.
\newblock Sparse models and methods for optimal instruments with an application
  to eminent domain.
\newblock \emph{Econometrica}, 80\penalty0 (6):\penalty0 2369--2429, 2012.

\bibitem[Zhao et~al.(2019)Zhao, Chen, Wang, and Small]{Zhao2019}
Zhao, Q., Chen, Y., Wang, J., and Small, D.~S.
\newblock {Powerful three-sample genome-wide design and robust statistical
  inference in summary-data Mendelian randomization}.
\newblock \emph{Int J Epidemiol}, 48\penalty0 (5):\penalty0 1478--1492, 07
  2019.

\bibitem[Grant and Burgess(2019)]{GB2019}
Grant, A.~J. and Burgess, S.
\newblock An efficient and robust approach to {M}endelian randomization with
  measured pleiotropic effects in a high-dimensional setting.
\newblock \emph{arXiv:1911.00347}, 2019.

\bibitem[Bi et~al.(2019)Bi, Kang, and Taylor]{Bi2019}
Bi, N., Kang, H., and Taylor, J.
\newblock Inference after selecting plausibly valid instruments with
  application to mendelian randomization.
\newblock \emph{arXiv:1911.03985}, 2019.

\bibitem[Davies et~al.(2019)Davies, Hill, Anderson, Sanderson,
  et~al.]{Davies2019}
Davies, N.~M., Hill, W.~D., Anderson, E.~L., Sanderson, E., et~al.
\newblock Multivariable two-sample {M}endelian randomization estimates of the
  effects of intelligence and education on health.
\newblock \emph{Elife}, 8:\penalty0 e43990, 2019.

\bibitem[Anderson et~al.(2020)Anderson, Howe, Wade, Ben-Shlomo,
  et~al.]{Anderson2020}
Anderson, E.~L., Howe, L.~D., Wade, K.~H., Ben-Shlomo, Y., et~al.
\newblock {Education, intelligence and {A}lzheimer’s disease: evidence from a
  multivariable two-sample {M}endelian randomization study}.
\newblock \emph{Int J Epidemiol}, 2020.
\newblock \doi{10.1093/ije/dyz280}.

\bibitem[Hill et~al.(2019)Hill, Marioni, Maghzian, Ritchie, et~al.]{Hill2019}
Hill, W.~D., Marioni, R.~E., Maghzian, O., Ritchie, S.~J., et~al.
\newblock {A combined analysis of genetically correlated traits identifies 187
  loci and a role for neurogenesis and myelination in intelligence}.
\newblock \emph{Mol. Psychiatry}, 24\penalty0 (2):\penalty0 169--181, 2019.

\bibitem[Okbay et~al.(2016)Okbay, Beauchamp, Fontana, Lee, et~al.]{Okbay2016}
Okbay, A., Beauchamp, J.~P., Fontana, M.~A., Lee, J.~J., et~al.
\newblock {Genome-wide association study identifies 74 loci associated with
  educational attainment}.
\newblock \emph{Nature}, 533:\penalty0 539--542, 2016.

\bibitem[Lambert et~al.(2013)Lambert, Ibrahim-Verbaas, Harold, Naj,
  et~al.]{Lambert2013}
Lambert, J.-C., Ibrahim-Verbaas, C.~A., Harold, D., Naj, A.~C., et~al.
\newblock {Meta-analysis of 74,046 individuals identifies 11 new susceptibility
  loci for {A}lzheimer's disease}.
\newblock \emph{Nat. Genet.}, 45:\penalty0 1452--1458, 2013.

\bibitem[Burgess et~al.(2016)Burgess, Dudbridge, and
  Thompson]{BDudbridge2016allele}
Burgess, S., Dudbridge, F., and Thompson, S.~G.
\newblock Combining information on multiple instrumental variables in
  {M}endelian randomization: comparison of allele score and summarized data
  methods.
\newblock \emph{Stat Med}, 35\penalty0 (11):\penalty0 1880--1906, 2016.

\bibitem[Sanderson et~al.(2020)Sanderson, Spiller, and Bowden]{Sanderson2020}
Sanderson, E., Spiller, W., and Bowden, J.
\newblock Testing and correcting for weak and pleiotropic instruments in
  two-sample multivariable {M}endelian randomisation.
\newblock \emph{bioRxiv}, 2020.
\newblock \doi{10.1101/2020.04.02.021980}.

\bibitem[Burgess and Thompson(2011)]{BurgessThompson2011}
Burgess, S. and Thompson, S.~G.
\newblock Bias in causal estimates from mendelian randomization studies with
  weak instruments.
\newblock \emph{Stat Med}, 30\penalty0 (11):\penalty0 1312--1323, 2011.

\bibitem[Yavorska and Staley(2019)]{mrpackage}
Yavorska, O. and Staley, J.
\newblock \emph{MendelianRandomization: Mendelian Randomization Package}, 2019.
\newblock URL \url{https://CRAN.R-project.org/package=MendelianRandomization}.
\newblock R package version 0.4.1.

\bibitem[Maechler et~al.(2020)Maechler, Rousseeuw, Croux, Todorov,
  et~al.]{robustbase}
Maechler, M., Rousseeuw, P., Croux, C., Todorov, V., et~al.
\newblock \emph{robustbase: Basic Robust Statistics}, 2020.
\newblock URL \url{http://robustbase.r-forge.r-project.org/}.
\newblock R package version 0.93-6.

\bibitem[Koenker(2020)]{quantreg}
Koenker, R.
\newblock \emph{quantreg: Quantile Regression}, 2020.
\newblock URL \url{https://CRAN.R-project.org/package=quantreg}.
\newblock R package version 5.55.

\bibitem[Friedman et~al.(2010)Friedman, Hastie, and Tibshirani]{glmnet}
Friedman, J., Hastie, T., and Tibshirani, R.
\newblock Regularization paths for generalized linear models via coordinate
  descent.
\newblock \emph{J Stat Softw}, 33\penalty0 (1):\penalty0 1--22, 2010.
\newblock URL \url{http://www.jstatsoft.org/v33/i01/}.

\end{thebibliography}

\section*{Supplementary material}

\setcounter{table}{0}
\setcounter{figure}{0}
\setcounter{equation}{0}
\renewcommand{\thesection}{S}
\renewcommand{\thetable}{S\arabic{table}}
\renewcommand{\thefigure}{S\arabic{figure}}

\subsection{Algorithm for performing the regularization approach} \label{se:lassoalg}
Let $\boldsymbol{\hat{\beta}_{X}}$ be the $p \times K$ matrix with ($j$, $k$)\textsuperscript{th} element $\hat{\beta}_{Xjk}$, $\boldsymbol{\hat{\beta}_{Y}}$ be the vector of length $p$ with $j$\textsuperscript{th} element $\hat{\beta}_{Yj}$, $\boldsymbol{S}$ be the $p \times p$ diagonal matrix with ($j$, $j$)\textsuperscript{th} element $\sigma_{Yj}^{-2}$ and $\boldsymbol{\theta_{0}}$ be the vector of length $p$ with $j$\textsuperscript{th} element $\theta_{0j}$. We denote by $\boldsymbol{P_{\hat{\beta}_{X}}} = \boldsymbol{S}^{1/2} \boldsymbol{\hat{\beta}_{X}} \left( \boldsymbol{\hat{\beta}_{X}'} \boldsymbol{S} \boldsymbol{\hat{\beta}_{X}} \right)^{-1} \boldsymbol{\hat{\beta}_{X}'} \boldsymbol{S}^{1/2}$ the projection onto the column space of $\boldsymbol{S}^{1/2} \boldsymbol{\hat{\beta}_{X}}$ and $\boldsymbol{I_{p}}$ the identity matrix of dimension $p$. We can solve (\ref{eq:mrlasso}), for a given value of $\lambda$, using the following procedure.
	\begin{enumerate}
		\item Let
		\[
		\boldsymbol{\hat{\theta}_{0\lambda}} = \argmin_{\theta_{0}} \Big\lVert \left( I_{p} - \boldsymbol{P_{\hat{\beta}_{X}}} \right) \boldsymbol{S}^{1/2} \left( \boldsymbol{\hat{\beta}_{Y}} - \boldsymbol{\theta_{0}} \right) \Big\rVert^{2} + \lambda \sum_{j=1}^{p} \left| \theta_{0j} \right| ,
		\]
		where $\lVert \cdot \rVert$ denotes the $\ell_{2}$ norm.
		\item let
		\[
		\boldsymbol{\hat{\theta}_{\lambda}} = \left( \boldsymbol{\hat{\beta}_{X}'} \boldsymbol{S} \boldsymbol{\hat{\beta}_{X}} \right)^{-1} \left( \boldsymbol{\hat{\beta}_{X}'} \boldsymbol{S} \boldsymbol{\hat{\beta}_{Y}} \right) .
		\]
	\end{enumerate}
The $k$\textsuperscript{th} element of $\boldsymbol{\hat{\theta}_{\lambda}}$ is the estimate of $\theta_{k}$ for given $\lambda$. Note that Step 1 is now a standard lasso, with responses $\left( I_{p} - \boldsymbol{P_{\hat{\beta}_{X}}} \right) \boldsymbol{S}^{1/2} \boldsymbol{\hat{\beta}_{Y}}$ and design matrix $\left( I_{p} - \boldsymbol{P_{\hat{\beta}_{X}}} \right) \boldsymbol{S}^{1/2}$, and can be computed with standard software.

\subsection{Supplementary simulation results} \label{se:supp_sims}
\begin{figure}
	\centering
	\includegraphics{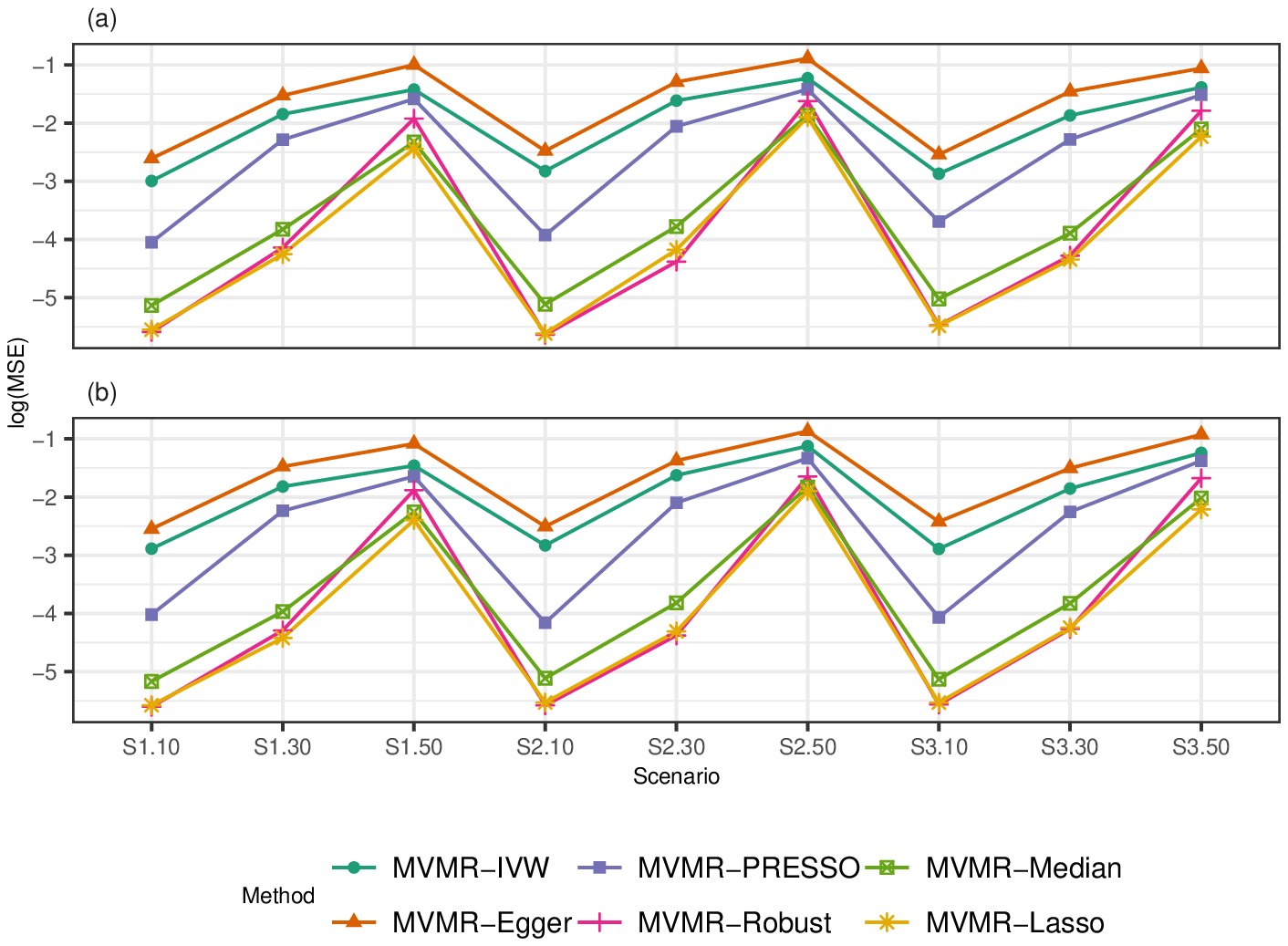}
	\caption{Logarithm of the mean squared errors for each scenario (S1, S2 and S3) and proportion of invalid genetic variants (10, 30 or 50\%), when $p=20$ and (a) $\theta_{1} = 0.2$ and (b) $\theta_{1} = 0$.}
	\label{fg:msep20}
\end{figure}

\begin{figure}
	\centering
	\includegraphics{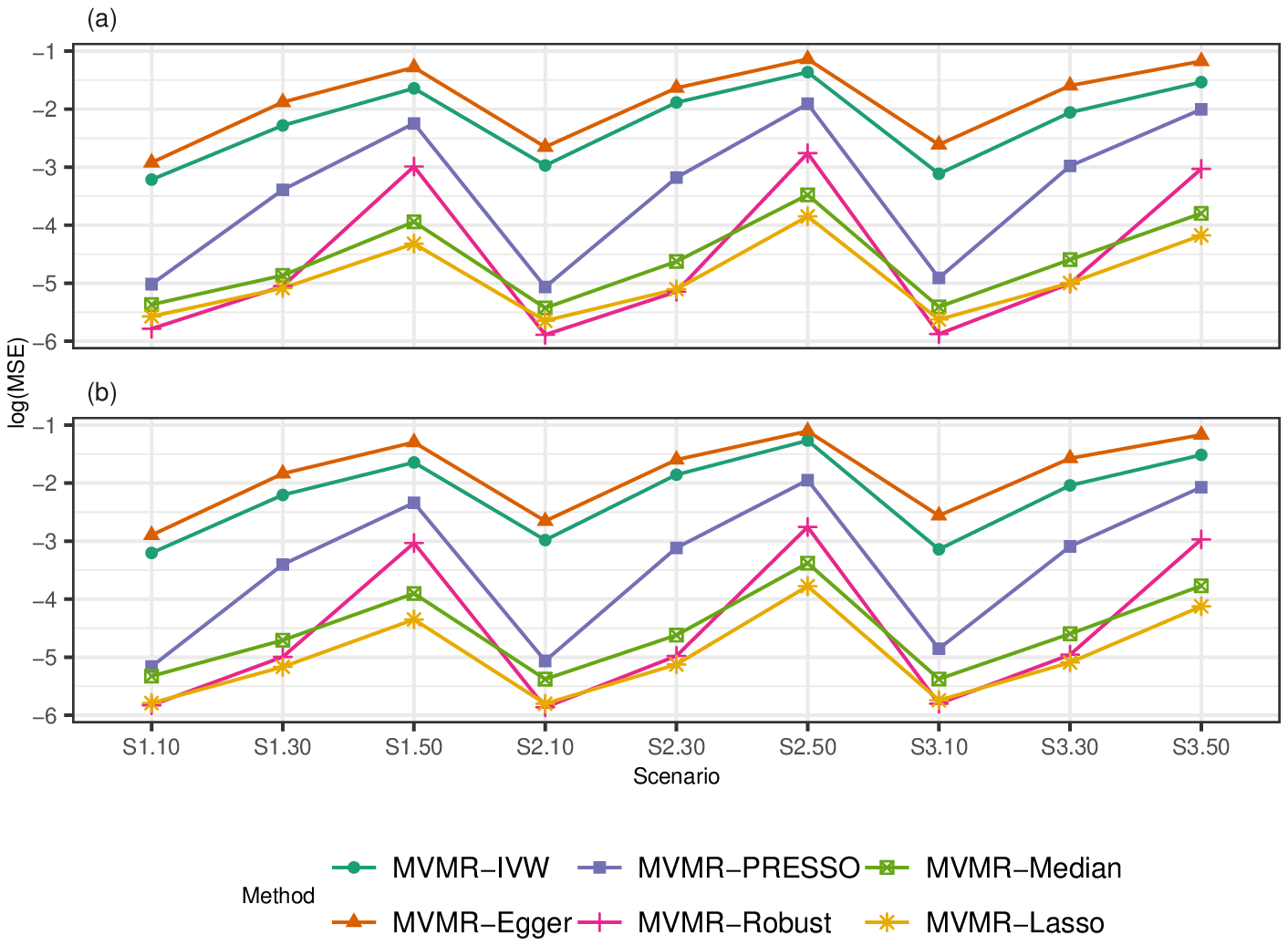}
	\caption{Logarithm of the mean squared errors for each scenario (S1, S2 and S3) and proportion of invalid genetic variants (10, 30 or 50\%), when the $\varepsilon_{Wij}$'s are correlated and (a) $\theta_{1} = 0.2$ and (b) $\theta_{1} = 0$.}
	\label{fg:msecor}
\end{figure}

\begin{figure}
	\centering
	\includegraphics{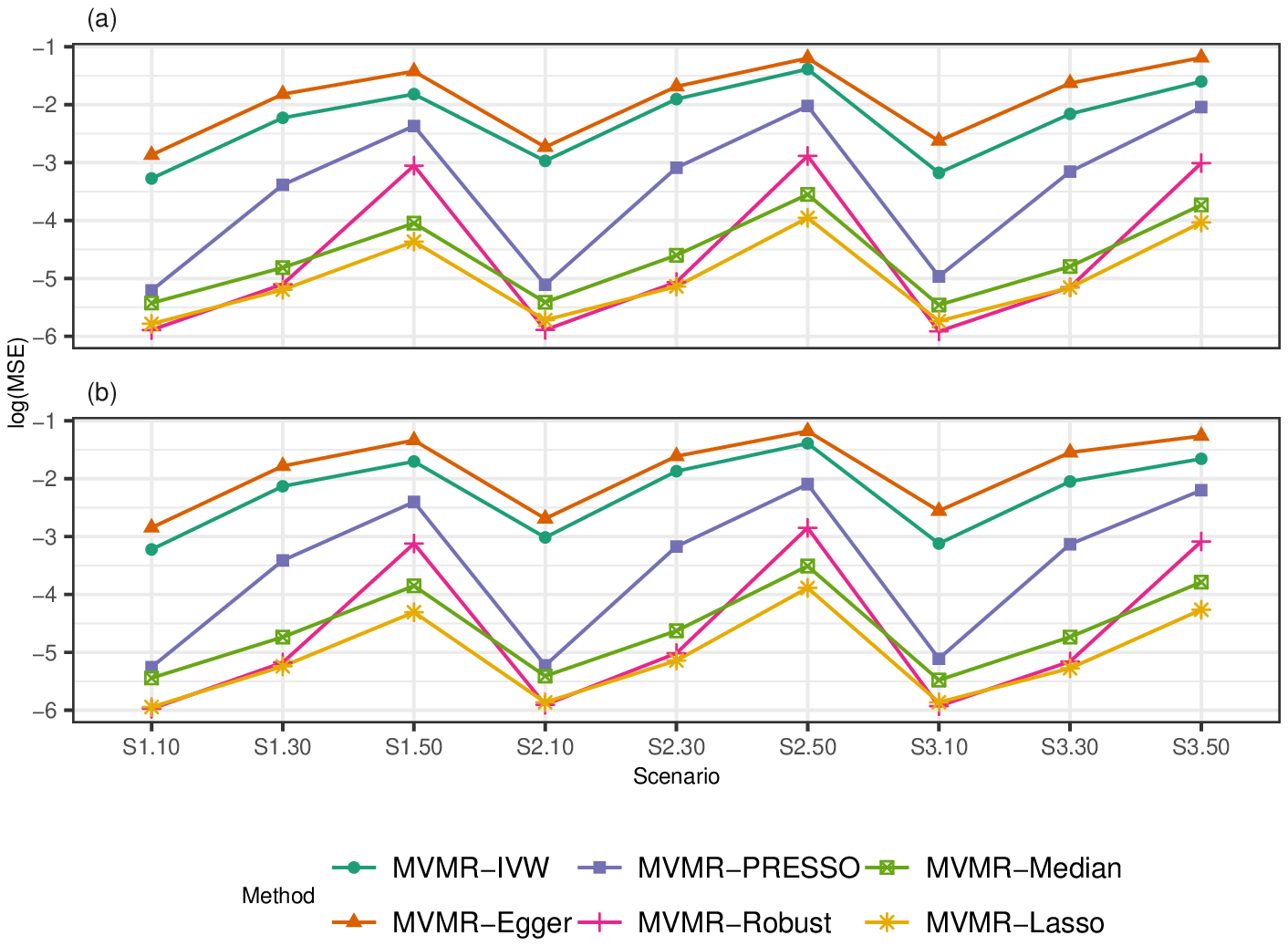}
	\caption{Logarithm of the mean squared errors for each scenario (S1, S2 and S3) and proportion of invalid genetic variants (10, 30 or 50\%), when the genetic variant-risk factor and genetic variant-outcome associations are estimated in the same sample and (a) $\theta_{1} = 0.2$ and (b) $\theta_{1} = 0$.}
	\label{fg:mseS1}
\end{figure}

In this section we present the results of the supplementary simulation studies described in Section \ref{se:sims}. Tables \ref{tb:simsHAp20}--\ref{tb:simsH0p20} and Figure \ref{fg:msep20} show the results for the case where $p=20$ and $\beta_{Xjk} \sim \textrm{Uniform} \left( 0, 0.22 \right)$. Tables \ref{tb:simsHAcor}--\ref{tb:simsH0cor} and Figure \ref{fg:msecor} show the results for the case where the risk factors are correlated with $\textrm{cor} \left( v_{Xik}, v_{Xil} \right) = 0.5$ for all $k \neq l$. Tables \ref{tb:simsHAS1}--\ref{tb:simsH0S1} and Figure \ref{fg:mseS1} show the results for the case where the genetic variant-trait associations were all estimated from the same sample (one sample Mendelian randomization).

\begin{figure}
	\centering
	\includegraphics{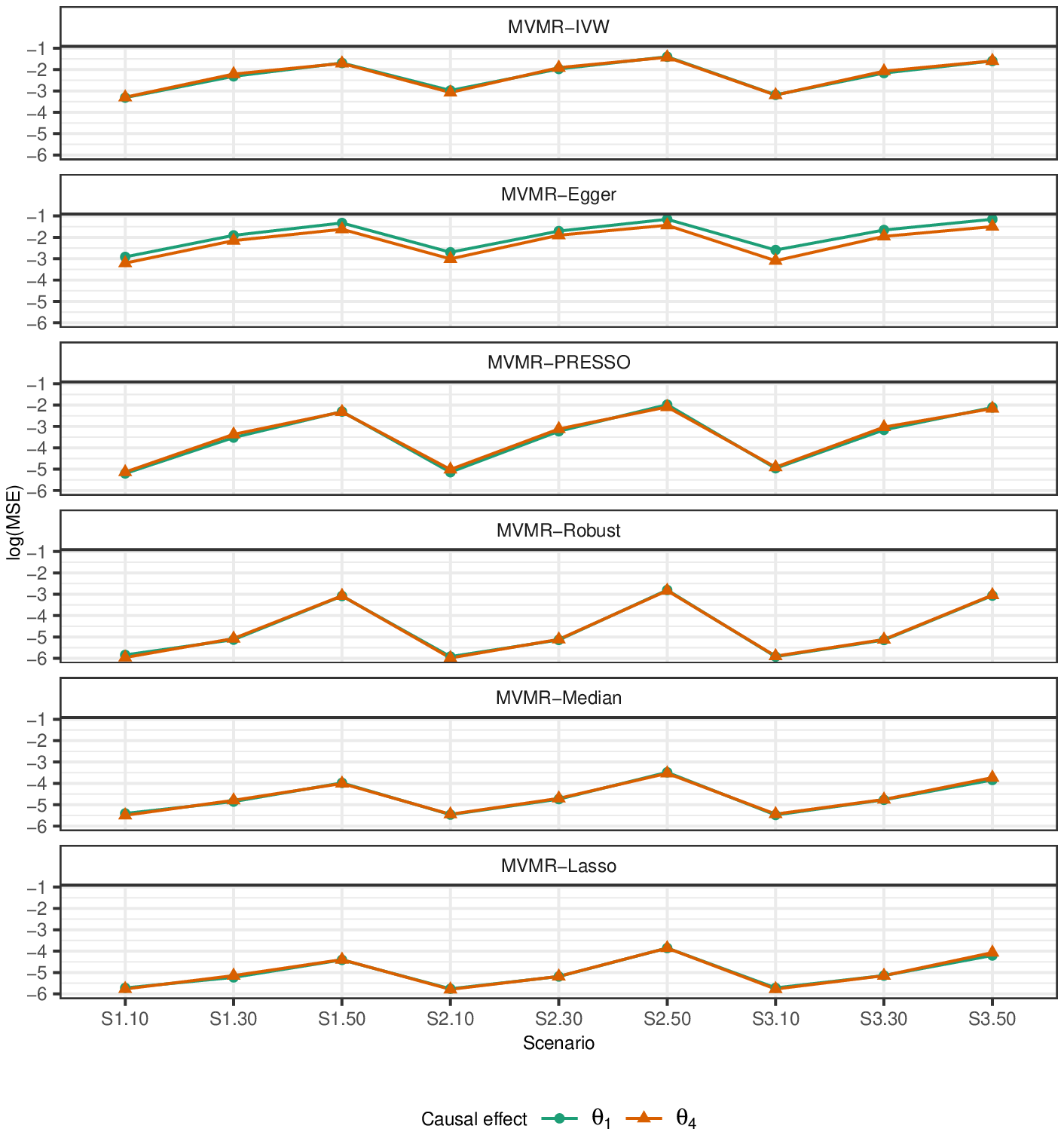}
	\caption{Logarithm of the mean squared errors for each scenario (S1, S2 and S3) and proportion of invalid genetic variants (10, 30 or 50\%), for the causal effect estimates for the first risk factor ($\theta_{1}$) and the fourth risk factor ($\theta_{4}$).}
	\label{fg:mse_th4}
\end{figure}

\begin{table}[p]
	\centering
	\caption{Mean and standard deviation (SD) of estimates, mean standard error of estimates (SE) and power when $p=20$ and $\theta_{1}=0.2$.}
	\label{tb:simsHAp20}
	\scalebox{0.8}{
		\begin{tabular}{l c c c c c c c c c c c c c c}
			\hline
			& \multicolumn{4}{c}{$10\%$ invalid} & & \multicolumn{4}{c}{$30\%$ invalid} & & \multicolumn{4}{c}{$50\%$ invalid} \\ \cline{2-5} \cline{7-10} \cline{12-15}
			Method & Mean & SD & SE & Power & & Mean & SD & SE &  Power & &Mean & SD & SE & Power \\ \hline
			& \multicolumn{14}{c}{Scenario 1: Balanced pleiotropy, InSIDE met} \\ \hline
			MVMR-IVW&0.198&0.224&0.203&0.280&&0.207&0.398&0.363&0.129&&0.188&0.491&0.474&0.084
			\\
			MVMR-Egger&0.206&0.272&0.241&0.241&&0.208&0.467&0.436&0.117&&0.183&0.607&0.569&0.086
			\\
			MVMR-PRESSO&0.200&0.132&0.103&0.602&&0.196&0.319&0.247&0.225&&0.195&0.453&0.369&0.148
			\\
			MVMR-Robust&0.201&0.061&0.067&0.831&&0.197&0.126&0.121&0.543&&0.206&0.382&0.406&0.172
			\\
			MVMR-Median&0.201&0.077&0.098&0.578&&0.197&0.148&0.124&0.402&&0.199&0.313&0.169&0.365
			\\
			MVMR-Lasso&0.201&0.062&0.072&0.832&&0.197&0.119&0.083&0.688&&0.207&0.295&0.112&0.589\\ \hline
			& \multicolumn{14}{c}{Scenario 2: Directional pleiotropy, InSIDE met} \\ \hline
			MVMR-IVW&0.217&0.243&0.221&0.271&&0.282&0.439&0.400&0.143&&0.283&0.535&0.504&0.111
			\\
			MVMR-Egger&0.197&0.290&0.264&0.214&&0.228&0.523&0.479&0.103&&0.200&0.642&0.603&0.095
			\\
			MVMR-PRESSO&0.205&0.141&0.102&0.619&&0.248&0.355&0.273&0.212&&0.269&0.487&0.398&0.153
			\\
			MVMR-Robust&0.197&0.060&0.065&0.831&&0.208&0.112&0.121&0.574&&0.242&0.442&0.446&0.144
			\\
			MVMR-Median&0.200&0.078&0.099&0.558&&0.216&0.151&0.129&0.430&&0.235&0.392&0.191&0.373
			\\
			MVMR-Lasso&0.198&0.060&0.073&0.818&&0.207&0.124&0.085&0.700&&0.230&0.386&0.123&0.544\\ \hline
			& \multicolumn{14}{c}{Scenario 3: Directional pleiotropy, InSIDE violated} \\ \hline
			MVMR-IVW&0.200&0.238&0.201&0.293&&0.234&0.391&0.373&0.125&&0.260&0.497&0.483&0.100
			\\
			MVMR-Egger&0.209&0.281&0.240&0.244&&0.248&0.480&0.445&0.115&&0.247&0.588&0.571&0.089
			\\
			MVMR-PRESSO&0.204&0.158&0.107&0.596&&0.227&0.319&0.267&0.201&&0.252&0.467&0.388&0.148
			\\
			MVMR-Robust&0.203&0.065&0.065&0.834&&0.195&0.118&0.121&0.552&&0.244&0.407&0.399&0.172
			\\
			MVMR-Median&0.205&0.081&0.099&0.587&&0.204&0.143&0.125&0.418&&0.240&0.348&0.180&0.362
			\\
			MVMR-Lasso&0.203&0.064&0.074&0.814&&0.199&0.114&0.084&0.689&&0.227&0.327&0.118&0.534\\ \hline
	\end{tabular}}
\end{table}

\begin{table}[p]
	\centering
	\caption{Mean and standard deviation (SD) of estimates, mean standard error of estimates (SE) and type I error rate when $p=20$ and $\theta_{1}=0$.}
	\label{tb:simsH0p20}
	\scalebox{0.8}{
		\begin{tabular}{l c c c c c c c c c c c c c c}
			\hline
			& \multicolumn{4}{c}{$10\%$ invalid} & & \multicolumn{4}{c}{$30\%$ invalid} & & \multicolumn{4}{c}{$50\%$ invalid} \\ \cline{2-5} \cline{7-10} \cline{12-15}
			Method & Mean & SD & SE & Type I & & Mean & SD & SE & Type I & &Mean & SD & SE & Type I \\ \hline
			& \multicolumn{14}{c}{Scenario 1: Balanced pleiotropy, InSIDE met} \\ \hline
			MVMR-IVW&0.014&0.236&0.205&0.068&&-0.006&0.403&0.366&0.077&&0.017&0.482&0.468&0.060
			\\
			MVMR-Egger&0.013&0.280&0.245&0.053&&-0.004&0.478&0.435&0.078&&0.020&0.582&0.563&0.069
			\\
			MVMR-PRESSO&0.006&0.134&0.099&0.073&&-0.001&0.327&0.246&0.090&&0.015&0.439&0.359&0.087
			\\
			MVMR-Robust&0.001&0.061&0.060&0.094&&0.003&0.117&0.120&0.086&&-0.002&0.391&0.396&0.084
			\\
			MVMR-Median&-0.001&0.076&0.084&0.025&&0.003&0.138&0.108&0.061&&0.007&0.324&0.155&0.196
			\\
			MVMR-Lasso&0.001&0.062&0.062&0.050&&-0.002&0.110&0.075&0.104&&0.007&0.302&0.103&0.281\\ \hline
			& \multicolumn{14}{c}{Scenario 2: Directional pleiotropy, InSIDE met} \\ \hline
			MVMR-IVW&0.018&0.242&0.219&0.062&&0.075&0.438&0.401&0.086&&0.091&0.563&0.507&0.094
			\\
			MVMR-Egger&0.001&0.286&0.263&0.056&&0.020&0.503&0.478&0.072&&-0.007&0.649&0.606&0.075\\
			MVMR-PRESSO&0.006&0.125&0.098&0.089&&0.060&0.345&0.267&0.094&&0.058&0.511&0.403&0.115
			\\
			MVMR-Robust&0.003&0.061&0.060&0.096&&0.006&0.112&0.119&0.093&&0.045&0.437&0.462&0.087
			\\
			MVMR-Median&0.006&0.077&0.084&0.037&&0.014&0.148&0.112&0.077&&0.033&0.400&0.177&0.252
			\\
			MVMR-Lasso&0.003&0.063&0.062&0.054&&0.007&0.116&0.076&0.108&&0.030&0.386&0.112&0.360\\ \hline
			& \multicolumn{14}{c}{Scenario 3: Directional pleiotropy, InSIDE violated} \\ \hline
			MVMR-IVW&0.017&0.235&0.212&0.070&&0.040&0.394&0.370&0.073&&0.072&0.533&0.487&0.078
			\\
			MVMR-Egger&0.033&0.296&0.253&0.075&&0.055&0.468&0.443&0.074&&0.057&0.627&0.580&0.075
			\\
			MVMR-PRESSO&0.011&0.130&0.103&0.082&&0.024&0.323&0.258&0.093&&0.067&0.498&0.392&0.100
			\\
			MVMR-Robust&0.001&0.062&0.060&0.111&&0.006&0.118&0.120&0.098&&0.039&0.431&0.432&0.093
			\\
			MVMR-Median&0.002&0.077&0.085&0.019&&0.007&0.148&0.110&0.077&&0.028&0.364&0.165&0.226
			\\
			MVMR-Lasso&0.001&0.063&0.064&0.051&&0.007&0.120&0.076&0.121&&0.016&0.331&0.106&0.315\\ \hline
	\end{tabular}}
\end{table}

\begin{table}[p]
	\centering
	\caption{Mean and standard deviation (SD) of estimates, mean standard error of estimates (SE) and power when the $v_{Xik}$'s are correlated and $\theta_{1}=0.2$.}
	\label{tb:simsHAcor}
	\scalebox{0.8}{
		\begin{tabular}{l c c c c c c c c c c c c c c}
			\hline
			& \multicolumn{4}{c}{$10\%$ invalid} & & \multicolumn{4}{c}{$30\%$ invalid} & & \multicolumn{4}{c}{$50\%$ invalid} \\ \cline{2-5} \cline{7-10} \cline{12-15}
			Method & Mean & SD & SE & Power & & Mean & SD & SE &  Power & &Mean & SD & SE & Power \\ \hline
			& \multicolumn{14}{c}{Scenario 1: Balanced pleiotropy, InSIDE met} \\ \hline
			MVMR-IVW&0.200&0.201&0.191&0.208&&0.200&0.320&0.331&0.090&&0.205&0.440&0.427&0.079
			\\
			MVMR-Egger&0.204&0.232&0.228&0.175&&0.217&0.391&0.394&0.092&&0.220&0.527&0.508&0.072
			\\
			MVMR-PRESSO&0.203&0.081&0.071&0.778&&0.197&0.184&0.151&0.336&&0.189&0.325&0.233&0.208
			\\
			MVMR-Robust&0.206&0.055&0.060&0.936&&0.202&0.080&0.081&0.708&&0.193&0.224&0.241&0.135
			\\
			MVMR-Median&0.205&0.068&0.088&0.681&&0.203&0.088&0.106&0.488&&0.200&0.139&0.135&0.335
			\\
			MVMR-Lasso&0.206&0.061&0.063&0.908&&0.204&0.079&0.073&0.793&&0.203&0.115&0.088&0.607\\ \hline
			& \multicolumn{14}{c}{Scenario 2: Directional pleiotropy, InSIDE met} \\ \hline
			MVMR-IVW&0.255&0.220&0.213&0.249&&0.335&0.366&0.360&0.159&&0.411&0.460&0.457&0.146
			\\
			MVMR-Egger&0.237&0.263&0.254&0.173&&0.260&0.438&0.429&0.105&&0.270&0.562&0.543&0.086
			\\
			MVMR-PRESSO&0.209&0.079&0.073&0.811&&0.242&0.200&0.167&0.369&&0.343&0.358&0.268&0.309
			\\
			MVMR-Robust&0.202&0.053&0.060&0.936&&0.200&0.076&0.082&0.701&&0.257&0.245&0.273&0.140
			\\
			MVMR-Median&0.205&0.066&0.089&0.680&&0.214&0.098&0.111&0.514&&0.246&0.170&0.151&0.387
			\\
			MVMR-Lasso&0.204&0.059&0.064&0.905&&0.204&0.078&0.076&0.772&&0.225&0.144&0.095&0.625\\ \hline
			& \multicolumn{14}{c}{Scenario 3: Directional pleiotropy, InSIDE violated} \\ \hline
			MVMR-IVW&0.240&0.207&0.197&0.256&&0.295&0.345&0.340&0.154&&0.314&0.450&0.433&0.114
			\\
			MVMR-Egger&0.273&0.261&0.233&0.257&&0.351&0.425&0.395&0.176&&0.369&0.530&0.499&0.127
			\\
			MVMR-PRESSO&0.208&0.086&0.077&0.774&&0.253&0.219&0.174&0.361&&0.285&0.357&0.271&0.260
			\\
			MVMR-Robust&0.203&0.053&0.060&0.941&&0.208&0.082&0.082&0.732&&0.235&0.217&0.248&0.142
			\\
			MVMR-Median&0.205&0.067&0.089&0.688&&0.219&0.099&0.109&0.523&&0.226&0.148&0.141&0.375
			\\
			MVMR-Lasso&0.204&0.060&0.064&0.913&&0.207&0.082&0.075&0.773&&0.217&0.123&0.091&0.628\\ \hline
	\end{tabular}}
\end{table}

\begin{table}[p]
	\centering
	\caption{Mean and standard deviation (SD) of estimates, mean standard error of estimates (SE) and type I error rate when the $v_{Xik}$'s are correlated and $\theta_{1}=0$.}
	\label{tb:simsH0cor}
	\scalebox{0.8}{
		\begin{tabular}{l c c c c c c c c c c c c c c}
			\hline
			& \multicolumn{4}{c}{$10\%$ invalid} & & \multicolumn{4}{c}{$30\%$ invalid} & & \multicolumn{4}{c}{$50\%$ invalid} \\ \cline{2-5} \cline{7-10} \cline{12-15}
			Method & Mean & SD & SE & Type I & & Mean & SD & SE & Type I & &Mean & SD & SE & Type I \\ \hline
			& \multicolumn{14}{c}{Scenario 1: Balanced pleiotropy, InSIDE met} \\ \hline
			MVMR-IVW&0.007&0.202&0.193&0.059&&0.009&0.332&0.332&0.043&&0.007&0.440&0.425&0.056
			\\
			MVMR-Egger&0.015&0.235&0.230&0.058&&0.013&0.399&0.396&0.054&&0.005&0.523&0.507&0.059
			\\
			MVMR-PRESSO&0.004&0.076&0.067&0.069&&0.003&0.183&0.142&0.092&&-0.002&0.311&0.225&0.112\\
			MVMR-Robust&0.003&0.054&0.053&0.056&&0.001&0.082&0.081&0.042&&-0.005&0.220&0.241&0.026
			\\
			MVMR-Median&0.004&0.070&0.074&0.040&&0.005&0.095&0.091&0.053&&-0.002&0.142&0.118&0.088
			\\
			MVMR-Lasso&0.003&0.055&0.053&0.060&&0.003&0.076&0.063&0.092&&-0.002&0.114&0.077&0.162\\ \hline
			& \multicolumn{14}{c}{Scenario 2: Directional pleiotropy, InSIDE met} \\ \hline
			MVMR-IVW&0.046&0.221&0.212&0.064&&0.119&0.377&0.362&0.074&&0.242&0.472&0.454&0.101
			\\
			MVMR-Egger&0.028&0.264&0.253&0.056&&0.044&0.448&0.431&0.066&&0.113&0.565&0.541&0.066
			\\
			MVMR-PRESSO&0.006&0.079&0.070&0.052&&0.042&0.206&0.163&0.089&&0.137&0.352&0.259&0.139
			\\
			MVMR-Robust&0.004&0.053&0.053&0.052&&0.003&0.083&0.082&0.050&&0.065&0.244&0.269&0.016
			\\
			MVMR-Median&0.006&0.068&0.074&0.024&&0.018&0.098&0.095&0.062&&0.053&0.177&0.132&0.137
			\\
			MVMR-Lasso&0.004&0.055&0.053&0.058&&0.004&0.077&0.065&0.108&&0.032&0.148&0.084&0.241\\ \hline
			& \multicolumn{14}{c}{Scenario 3: Directional pleiotropy, InSIDE violated} \\ \hline
			MVMR-IVW&0.047&0.203&0.196&0.055&&0.087&0.350&0.340&0.074&&0.132&0.451&0.432&0.070
			\\
			MVMR-Egger&0.082&0.266&0.232&0.089&&0.152&0.429&0.396&0.091&&0.177&0.529&0.499&0.079
			\\
			MVMR-PRESSO&0.014&0.087&0.073&0.070&&0.043&0.209&0.166&0.092&&0.079&0.346&0.262&0.108
			\\
			MVMR-Robust&0.007&0.055&0.053&0.052&&0.012&0.083&0.082&0.053&&0.029&0.225&0.248&0.015
			\\
			MVMR-Median&0.010&0.067&0.073&0.032&&0.019&0.099&0.093&0.065&&0.026&0.150&0.123&0.100
			\\
			MVMR-Lasso&0.007&0.056&0.052&0.057&&0.012&0.078&0.064&0.106&&0.015&0.126&0.080&0.196\\ \hline
	\end{tabular}}
\end{table}

\begin{table}[p]
	\centering
	\caption{Mean and standard deviation (SD) of estimates, mean standard error of estimates (SE) and power when the genetic variant-risk factor and genetic variant-outcome associations are estimated in the same sample and $\theta_{1}=0.2$.}
	\label{tb:simsHAS1}
	\scalebox{0.8}{
		\begin{tabular}{l c c c c c c c c c c c c c c}
			\hline
			& \multicolumn{4}{c}{$10\%$ invalid} & & \multicolumn{4}{c}{$30\%$ invalid} & & \multicolumn{4}{c}{$50\%$ invalid} \\ \cline{2-5} \cline{7-10} \cline{12-15}
			Method & Mean & SD & SE & Power & & Mean & SD & SE &  Power & &Mean & SD & SE & Power \\ \hline
			& \multicolumn{14}{c}{Scenario 1: Balanced pleiotropy, InSIDE met} \\ \hline
			MVMR-IVW&0.209&0.194&0.185&0.231&&0.203&0.328&0.320&0.112&&0.213&0.403&0.413&0.078
			\\
			MVMR-Egger&0.226&0.237&0.225&0.210&&0.215&0.403&0.390&0.105&&0.234&0.489&0.502&0.071
			\\
			MVMR-PRESSO&0.202&0.074&0.068&0.831&&0.197&0.184&0.138&0.392&&0.198&0.306&0.220&0.248
			\\
			MVMR-Robust&0.203&0.053&0.055&0.960&&0.199&0.078&0.078&0.726&&0.191&0.217&0.230&0.159
			\\
			MVMR-Median&0.203&0.066&0.082&0.752&&0.198&0.090&0.100&0.528&&0.200&0.132&0.126&0.374
			\\
			MVMR-Lasso&0.203&0.055&0.058&0.945&&0.199&0.074&0.068&0.801&&0.197&0.113&0.082&0.636\\ \hline
			& \multicolumn{14}{c}{Scenario 2: Directional pleiotropy, InSIDE met} \\ \hline
			MVMR-IVW&0.263&0.217&0.207&0.282&&0.346&0.357&0.351&0.181&&0.438&0.440&0.441&0.171
			\\
			MVMR-Egger&0.248&0.251&0.251&0.186&&0.272&0.424&0.426&0.096&&0.305&0.540&0.534&0.102
			\\
			MVMR-PRESSO&0.206&0.077&0.069&0.809&&0.257&0.206&0.159&0.442&&0.337&0.337&0.256&0.337
			\\
			MVMR-Robust&0.202&0.053&0.056&0.950&&0.206&0.079&0.079&0.745&&0.252&0.231&0.257&0.146
			\\
			MVMR-Median&0.206&0.067&0.083&0.737&&0.222&0.098&0.105&0.569&&0.247&0.163&0.142&0.435
			\\
			MVMR-Lasso&0.203&0.057&0.059&0.923&&0.207&0.076&0.071&0.819&&0.227&0.136&0.089&0.672\\ \hline
			& \multicolumn{14}{c}{Scenario 3: Directional pleiotropy, InSIDE violated} \\ \hline
			MVMR-IVW&0.235&0.201&0.192&0.262&&0.282&0.329&0.329&0.138&&0.324&0.432&0.422&0.119
			\\
			MVMR-Egger&0.275&0.259&0.231&0.256&&0.350&0.417&0.390&0.175&&0.392&0.518&0.494&0.142
			\\
			MVMR-PRESSO&0.209&0.083&0.074&0.799&&0.241&0.202&0.162&0.403&&0.285&0.350&0.254&0.284
			\\
			MVMR-Robust&0.203&0.052&0.055&0.961&&0.208&0.076&0.079&0.766&&0.233&0.220&0.250&0.132
			\\
			MVMR-Median&0.205&0.065&0.083&0.745&&0.219&0.089&0.102&0.592&&0.231&0.152&0.135&0.420
			\\
			MVMR-Lasso&0.205&0.057&0.059&0.939&&0.210&0.075&0.070&0.836&&0.222&0.131&0.086&0.661\\ \hline
	\end{tabular}}
\end{table}

\begin{table}[p]
	\centering
	\caption{Mean and standard deviation (SD) of estimates, mean standard error of estimates (SE) and power when the genetic variant-risk factor and genetic variant-outcome associations are estimated in the same sample and $\theta_{1}=0$.}
	\label{tb:simsH0S1}
	\scalebox{0.8}{
		\begin{tabular}{l c c c c c c c c c c c c c c}
			\hline
			& \multicolumn{4}{c}{$10\%$ invalid} & & \multicolumn{4}{c}{$30\%$ invalid} & & \multicolumn{4}{c}{$50\%$ invalid} \\ \cline{2-5} \cline{7-10} \cline{12-15}
			Method & Mean & SD & SE & Type I & & Mean & SD & SE & Type I & &Mean & SD & SE & Type I \\ \hline
			& \multicolumn{14}{c}{Scenario 1: Balanced pleiotropy, InSIDE met} \\ \hline
			MVMR-IVW&0.000&0.200&0.188&0.066&&0.001&0.345&0.323&0.057&&-0.007&0.427&0.414&0.059
			\\
			MVMR-Egger&0.018&0.240&0.229&0.058&&-0.006&0.410&0.392&0.054&&0.009&0.513&0.503&0.049
			\\
			MVMR-PRESSO&-0.003&0.072&0.065&0.053&&0.002&0.182&0.136&0.079&&-0.001&0.301&0.216&0.105
			\\
			MVMR-Robust&0.003&0.050&0.052&0.047&&0.002&0.075&0.078&0.051&&0.006&0.210&0.235&0.018
			\\
			MVMR-Median&0.003&0.066&0.072&0.031&&0.003&0.094&0.089&0.052&&-0.001&0.146&0.115&0.117
			\\
			MVMR-Lasso&0.003&0.051&0.051&0.051&&0.001&0.073&0.061&0.092&&0.002&0.116&0.076&0.183\\ \hline
			& \multicolumn{14}{c}{Scenario 2: Directional pleiotropy, InSIDE met} \\ \hline
			MVMR-IVW&0.053&0.215&0.206&0.067&&0.140&0.366&0.350&0.075&&0.248&0.433&0.441&0.079
			\\
			MVMR-Egger&0.037&0.258&0.250&0.051&&0.087&0.438&0.426&0.054&&0.117&0.542&0.535&0.066
			\\
			MVMR-PRESSO&0.007&0.073&0.066&0.063&&0.045&0.200&0.152&0.090&&0.131&0.326&0.244&0.147
			\\
			MVMR-Robust&0.003&0.052&0.052&0.061&&0.010&0.081&0.080&0.048&&0.058&0.234&0.261&0.027
			\\
			MVMR-Median&0.006&0.067&0.072&0.035&&0.020&0.097&0.092&0.061&&0.052&0.165&0.128&0.121
			\\
			MVMR-Lasso&0.003&0.053&0.052&0.056&&0.011&0.076&0.063&0.108&&0.033&0.139&0.081&0.242\\ \hline
			& \multicolumn{14}{c}{Scenario 3: Directional pleiotropy, InSIDE violated} \\ \hline
			MVMR-IVW&0.028&0.208&0.192&0.059&&0.108&0.343&0.328&0.069&&0.119&0.420&0.418&0.063
			\\
			MVMR-Egger&0.067&0.270&0.230&0.097&&0.178&0.426&0.390&0.097&&0.179&0.501&0.490&0.071
			\\
			MVMR-PRESSO&0.007&0.078&0.071&0.061&&0.050&0.203&0.156&0.092&&0.077&0.324&0.248&0.119
			\\
			MVMR-Robust&0.002&0.052&0.052&0.059&&0.003&0.076&0.079&0.049&&0.023&0.213&0.242&0.031
			\\
			MVMR-Median&0.002&0.065&0.072&0.021&&0.016&0.092&0.090&0.056&&0.030&0.147&0.119&0.110
			\\
			MVMR-Lasso&0.002&0.053&0.051&0.067&&0.006&0.071&0.062&0.092&&0.016&0.118&0.077&0.191\\ \hline
	\end{tabular}}
\end{table}

\end{document}